\documentclass[pra,10pt,twocolumn]{revtex4}
\usepackage[dvips]{graphicx}
\usepackage{float}
\usepackage{amsfonts}
\usepackage{amsmath}
\usepackage{amssymb}
\usepackage{enumerate}
\usepackage{color}

\usepackage{hhline}
\usepackage{threeparttable}
\usepackage{supertabular}
\usepackage{multirow}
\usepackage{tabularx}
\usepackage{pslatex}

\begin{document}

\title{Investigations of a coherently driven semiconductor optical cavity QED system}

\author{Kartik Srinivasan}
\email{kartik.srinivasan@nist.gov} \affiliation{Center for
Nanoscale Science and Technology, National Institute of Standards
and Technology, Gaithersburg, MD 20899}
\author{Christopher P. Michael}
\author{Raviv Perahia}
\author{Oskar Painter}
\affiliation{Department of Applied Physics, California Institute
of Technology, Pasadena, California 91125}
\date{\today}
\begin{abstract} Chip-based cavity quantum electrodynamics (QED)
  devices consisting of a self-assembled InAs quantum dot (QD) coupled
  to a high quality factor GaAs microdisk cavity are coherently probed
  through their optical channel using a fiber taper waveguide.  We
  highlight one particularly important aspect of this all-fiber
  measurement setup, which is the accuracy to which the optical
  coupling level and optical losses are known relative to typical
  free-space excitation techniques. This allows for precise knowledge
  of the intracavity photon number and measurement of absolute
  transmitted and reflected signals.  Resonant optical spectroscopy of
  the system under both weak and strong driving conditions are
  presented, which when compared with a quantum master equation model
  of the system allows for determination of the coherent coupling rate
  between QD exciton and optical cavity mode, the different levels of
  elastic and inelastic dephasing of the exciton state, and the
  position and orientation of the QD within the cavity.  Pump-probe
  measurements are also performed in which a far off-resonant
  red-detuned control laser beam is introduced into the cavity.
  Rather than producing a measurable ac-Stark shift in the exciton
  line of the QD, we find that this control beam induces a saturation
  of the resonant system response.  The broad photoluminescence
  spectrum resulting from the presence of the control beam in the
  cavity points to sub-bandgap absorption in the semiconductor, and
  the resulting free-carrier generation, as the likely source of
  system saturation.

\end{abstract}
 \maketitle

\setcounter{page}{1}
\section{Introduction}
\label{sec:intro}

In recent years, experimental investigations of cavity quantum
electrodynamics (QED) have diversified from systems incorporating
cooled alkali atoms in high-finesse Fabry-Perot cavities
\cite{ref:Kimble2,ref:Hood,ref:Pinske} and Rydberg atoms in
superconducting microwave cavities \cite{ref:Raimond,ref:Walther}
to chip-based implementations involving Cooper pair boxes and
transmission line resonators \cite{ref:Walraff}, trapped atoms and
monolithic dielectric microcavities \cite{ref:Aoki1}, and
epitaxially grown quantum dots embedded in semiconductor optical
microcavities \cite{ref:Reithmaier,ref:Yoshie3,ref:Peter}.
Investigations of these new systems can proceed along a number of
different paths. At a first level, they seek to confirm the
observation of basic features of the Jaynes-Cummings model
\cite{ref:Jaynes_Cummings} that describes the interaction of a
two-level system with a quantized electromagnetic field.  At the
same time, phenomena specific to the experimental system at hand,
such as the influence of electron-phonon interactions in
semiconductors, are an important line of investigation. Thirdly,
the scalability of these chip-based architectures, resulting from
the planar fabrication techniques by which they are created, has
inspired numerous theoretical proposals for quantum information
processing \cite{ref:Imamoglu,ref:Blais2}, and recently,
experimental progress in the form of coupling of two
superconducting qubits through a resonant cavity
\cite{ref:Sillanpaa,ref:Majer}.

In this work, we study the interaction of a self-assembled InAs
quantum dot (QD) coupled to the optical mode of a high quality factor
($Q$) GaAs microdisk cavity, and present results that are largely
focused on addressing the first line of investigation described
above, but also touch upon the second.  In regards to the former,
we build upon recent experimental results \cite{ref:Srinivasan16},
where the cavity-QD system is $\textit{coherently}$ excited and
probed by a resonant optical field through use of a fiber taper
waveguide.  In comparison to the vast majority of semiconductor QD
cavity QED work
 \cite{ref:Reithmaier,ref:Yoshie3,ref:Peter,ref:Hennessy3,ref:Press},
which rely upon $\textit{incoherent}$ excitation and probing via
photoluminescence, the use of a resonant probe in this work (and
in the work of Ref. [\onlinecite{ref:Englund2}]) to address the
system through its optical channel provides a more analogous
correspondence with spectral measurements done in atomic cavity
QED \cite{ref:Hood2,ref:Boca}, and is a necessity for maintaining
the coherence required in certain quantum information processing
applications \cite{ref:Kiraz}.  Use of the fiber taper waveguide
coupling technique allows for an accurate estimate of quantities
such as the intracavity photon number and absolute transmitted and
reflected signals, due to the accuracy to which the
cavity-waveguide coupling efficiency and any losses in this
all-fiber measurement setup are known.  We present measurements of
the cavity's spectral response as it is tuned with respect to the
neutral exciton line of an isolated QD while operating in the
linear (weak driving) regime with an average intracavity photon
number $n_{\text{cav}}\ll1$. These measurements show clear
anti-crossing and vacuum Rabi splitting behavior with the system
operating in the strong coupling regime, where the single photon
cavity-QD coupling rate $g$ exceeds both the cavity loss rate
$\kappa$ and QD decay rate $\gamma_{\perp}$ \cite{ref:Kimble2}.
Fixing the cavity detuning with respect to the neutral exciton
transition of the QD, we also measure the spectral response as a
function of the probe beam power, and observe system saturation
for $n_{\text{cav}}\approx0.1$. A previously developed quantum
master equation model for this system \cite{ref:Srinivasan13} is
fit to the measured data, providing estimates for the cavity-QD
coupling strength, relative position and orientation of the QD
within the cavity, and exciton dephasing rates (both elastic and
inelastic dephasing).

The fact that the QD is embedded in a host semiconductor matrix
can give rise to a number of effects not typically seen in atomic
systems, such as spectral diffusion and phase destroying
collisional processes due to electron-phonon and electron-electron
scattering.  At the end of this paper, we consider the influence
of material absorption and subsequent free-carrier generation on
the saturation behavior of the coupled microdisk-QD system.  This
is done by characterizing the system in a pump-probe experiment.
By fixing the probe beam power to the weak driving limit
($n_{\text{cav}}\ll1$) and sweeping its frequency to trace out the
system's resonant spectral response, and then coupling a control
laser beam into a far red-detuned cavity mode, we can examine the
effects of below bandgap (in energy) excitation on the system. We
observe saturation for control beam powers that generate
$n_{\text{cav}}\approx1$ in the off-resonant cavity mode, about
one order of magnitude larger than the saturation $n_{\text{cav}}$
we observe when varying the probe beam power.  We attribute this
new saturation to absorption from bulk defect and surface states
of the semiconductor, and the subsequent generation of free
carrier charges that cause spectral diffusion, blinking, and
dephasing of the exciton states within the QD \cite{ref:Santori4}.
These processes may be of importance in future pump-probe
measurements of semiconductor cavity-QD systems, which could be
used to further investigate the structure of the Jaynes-Cummings
system \cite{ref:Thompson2,ref:Walraff2} or to provide a level of
control on the cavity-QD interaction \cite{ref:Schuster}, for
example, through the ac-Stark effect.

The organization of this paper is as follows: in section
\ref{sec:experimental_system}, we review the experimental system
studied and highlight some of the important aspects of the methods
used.  Section \ref{sec:device_preparation} describes the
fabrication and initial measurements done to identify suitable
devices for the coherent probing experiments.  Section
\ref{sec:strong_coupling_data} is an extension of the work
presented in Ref. [\onlinecite{ref:Srinivasan16}], and in
particular, presents some additional data on a strongly coupled
microdisk-QD system, as well as a more extended discussion of how
the experimental data was fit to the numerical model presented in
Ref. [\onlinecite{ref:Srinivasan13}]. Section
\ref{sec:bad_cavity_data} presents similar data, but for a system
that is in the bad cavity limit with $\kappa>g>\gamma_{\perp}$ and
$g\gtrsim(\kappa+\gamma_{\perp})/2$ (allowing for resolved vacuum
Rabi splitting in cavity transmission and reflection). At the end
of this section, we present data on the two saturation mechanisms
mentioned above, and discuss the possible origin of the resonant
system response saturation effected by the far off-resonance
control laser beam.

\section{Experimental system}
\label{sec:experimental_system}

%This section presents some important details related to the
%experiments, including a brief summary of fiber taper coupling to
%the microdisk cavities and a description of the experimental setup
%and tools used.

\subsection{Fiber taper waveguide coupling}

Two of the primary difficulties in performing resonant optical
measurements on the microcavity-QD system are effectively coupling
light into the system and then separating QD fluorescence from
incident light that has been scattered off defects, other portions
of the sample, etc. Recent experiments involving resonant
excitation of a QD \cite{ref:Muller,ref:Melet} have utilized
geometries in which the QD is excited by an in-plane beam and the
vertically emitted fluorescence is measured. Alternately, in Ref.
 [\onlinecite{ref:Englund2}], the authors direct a micron-scale probe beam
onto the cavity surface and measure the reflected signal in
cross-polarization, with an estimated coupling efficiency into the
cavity mode of 1 $\%$ to 2 $\%$.

\begin{figure}[t]
\begin{center}
\includegraphics[width=\columnwidth]{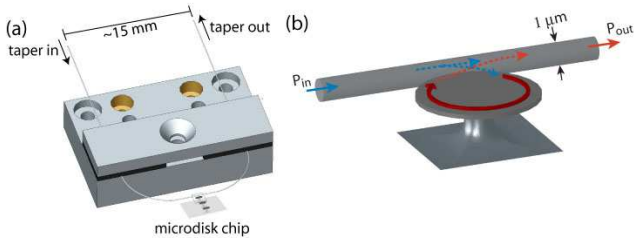}
\caption{Schematics of the optical fiber taper waveguide coupling
approach: (a) Aluminum mount used to hold the fiber taper and
position it with respect to an array of microdisk cavities; (b)
Zoomed-in region of the taper-cavity coupling
region.}\label{fig:taper_cav_coupling}
\end{center}
\end{figure}

The approach we use relies upon evanescent coupling between a
waveguide probe and the microcavity \cite{ref:Srinivasan16}.  The
waveguide probe is an optical fiber taper \cite{ref:Knight}, which
is a standard single mode fiber that has been heated and stretched
down to a wavelength-scale minimum diameter in a symmetric
fashion, so that the overall structure is a double-ended device in
which the input and output regions are both composed of standard
single mode fiber.  In the region of minimum fiber diameter, the
evanescent field of the waveguide mode extends significantly into
the surrounding air, so that it can excite the microcavity's modes
when it is brought close to it (Figure
\ref{fig:taper_cav_coupling}). The tapering process is done
adiabatically over a distance of 5 mm to 10 mm, so that its
transmission level is typically better than 50 $\%$ (fractional
transmission $\zeta=P_{\text{out}}/P_{\text{in}}=0.5$), and can
routinely be better than 90 $\%$ ($\zeta=$0.9).  For a properly
chosen cavity geometry and taper size, the regimes of critical
coupling and overcoupling can be achieved with little parasitic
coupling loss, even for a cavity fabricated in a high refractive
index semiconductor \cite{ref:Borselli_thesis}. The overall
coupling efficiency into the cavity mode, $\sqrt{\zeta}{\Delta}T$,
can exceed 10 $\%$ to 20 $\%$ for typical devices, where
${\Delta}T$ is the transmission contrast of the resonance dip in
the taper's transmission spectrum that results from coupling to
the cavity mode (${\Delta}T=1-T$, where $T$ is the normalized
transmission level on-resonance). This overall coupling efficiency
relates the dropped (coupled) power into the cavity
($P_{\text{d}}$) with the input power into the taper
($P_{\text{in}}$), with
$P_{\text{d}}=\sqrt{\zeta}{\Delta}T{P_{\text{in}}}$.

Perhaps more important than the magnitude of the coupling
efficiency is the accuracy to which the intracavity photon number
can be estimated for a given optical input power.  The fiber taper
waveguide allows one to probe the cavity-QD system through a
single electromagnetic spatial mode (there are actually two
degenerate taper modes of distinct polarization; however, by using
a fiber polarization controller one can select for the
polarization that matches the cavity mode of interest).  As the
fiber tapers are formed from single-mode optical fiber, the input,
transmission, and reflection signals carried by the fiber taper
are naturally filtered for the fundamental taper mode (all other
higher-order modes radiate away into the cladding of the
single-mode fiber).  For a single spatial mode, the transmission
contrast of the fiber-coupled cavity, a result of interference
between the light transmitted past the cavity with that coupled
into and re-emitted from the cavity, can be used along with the
width of the cavity spectrum to determine the coupling rate of
input light from the taper waveguide into the cavity mode and the
average intracavity photon number with high accuracy.  This should
be contrasted with a multi-mode excitation and detection scheme
(such as many free-space techniques) in which interference is only
partial and the determination of the actual intracavity photon
number is compromised.  The use of an all-fiber setup, where
system losses can be easily measured, also means that absolute
measurements of the transmitted and reflected signals can be made.
This allows us to quantitatively compare our experimental results
with a quantum master equation model for the system.  This is of
particular importance in trying to identify non-ideal behavior in
the experimental system and potential ways in which the QD behaves
differently than an ideal two-level system.

As discussed above, for a single spatial mode interacting with the
cavity, the cavity's transmission spectrum, along with knowledge of
$P_{\text{in}}$ and $\zeta$, allows us to determine the
\emph{on-resonance} internal cavity energy ($U$) and average
intracavity photon number ($n_{\text{cav}}$) through the relation,

\begin{equation}
\label{eq:intracavity_photon_number}
U(\omega_{0})=n_{\text{cav}}\hbar\omega_{0}=\sqrt{\zeta}{\Delta}TQ_{i+P}\left(\frac{P_{\text{in}}}{\omega_{0}}\right),
\end{equation}

\noindent where $\omega_{0}$ is the cavity mode resonance
frequency and $Q_{i+P}$ is the cavity $Q$ due to intrinsic losses
such as absorption and scattering as well as parasitic losses
stemming from cavity radiation into higher-order modes of the
fiber taper \cite{ref:Barclay7}.  Here we assume symmetric loss in
the taper (usually a very good approximation), so that
$\sqrt{\zeta}$ is the fractional transmission between taper input
and the taper-cavity coupling region. In practice, we don't
measure $Q_{i+P}$, but instead measure the total loaded quality
factor $Q_{T}$ (or more precisely a loaded linewidth
$\delta\omega_{T}=\omega_{0}/Q_{T}$), which in addition to
intrinsic and parasitic losses, also includes cavity radiation
into the fundamental mode of the fiber taper waveguide.
Off-resonance excitation of the cavity simply results in a cavity
energy which is scaled by a Lorentzian term, $1/(1 +
(\Delta/2\delta\omega_{T})^2)$, where $\Delta=\omega-\omega_{0}$
is the detuning.  Following the waveguide-cavity coupled mode
theory analysis presented in several other works
\cite{ref:Manolatou,ref:Spillane2,ref:Barclay7}, we can write

\begin{equation}
\label{eq:Q_i_P_formula}
Q_{i+P}=\frac{2Q_{T}}{1\pm\sqrt{1-{\Delta}T}},
\end{equation}

\noindent where $+$ ($-$) is appropriate for the undercoupled
(overcoupled) regime, in which the waveguide-cavity coupling rate
is less than (greater than) the total of the intrinsic and
parasitic loss rates. Replacing $Q_{i+P}$ in eq.
(\ref{eq:intracavity_photon_number}), we arrive at

\begin{equation}
\label{eq:intracavity_photon_number_2}
U(\omega_{0})=n_{\text{cav}}\hbar\omega_{0}=\frac{2\sqrt{\zeta}{\Delta}T P_{\text{in}} Q_{T}}{\bigl(1\pm\sqrt{1-\Delta{T}}\bigr)\omega_{0}},
\end{equation}

\noindent so that $U$ and $n_{\text{cav}}$ are written in terms of
the measured quantities $\omega_{0}$, $P_{\text{in}}$, $\zeta$,
${\Delta}T$, and $Q_{T}$.

Past measurements have shown that the fiber taper coupling method
can be applied to both semiconductor microdisks
\cite{ref:Borselli_thesis,ref:Srinivasan9} and photonic crystals
\cite{ref:Srinivasan7,ref:Barclay7}. We have chosen to focus our
cavity QED studies on microdisks due to the ability to achieve
efficient coupling without the necessity of an intermediate
on-chip waveguide element, as was used in Ref.
[\onlinecite{ref:Barclay7}], and the relative ease in cavity
fabrication. As described elsewhere \cite{ref:Painter_OE}, direct
taper coupling to $L3$ photonic crystal cavities
\cite{ref:Akahane2} can produce ${\Delta}T$ values in excess of 50
$\%$, but at such coupling levels a significant amount of
parasitic loss is also present, degrading the $Q$ from its
intrinsic level $Q_{i}=3{\times}10^4$ to a loaded level
$Q_{T}=1{\times}10^4$. In contrast, we have been able to directly
taper couple to ultra-small mode volume ($V_{\text{eff}}$)
microdisk cavities while achieving $Q_{T}=10^5$ and ${\Delta}T=60$
$\%$ \cite{ref:Srinivasan9}. Nevertheless, the ability to
investigate different cavity geometries is one strength of this
technique.  A second is its applicability to other solid-state
systems and materials of both high and low refractive index
\cite{ref:Barclay8}.

\subsection{Experimental setup}

\begin{figure}[t]
\begin{center}
\includegraphics[width=\columnwidth]{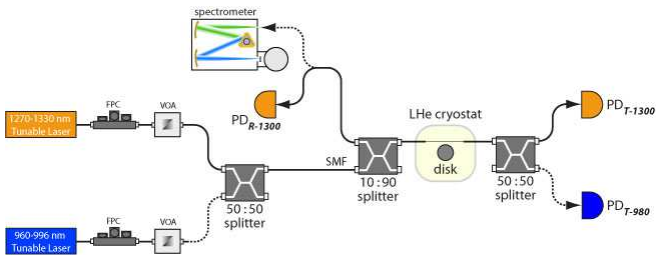}
\caption{Schematic of the experimental setup used for coherent
optical probing with a 1300 nm tunable laser (solid lines) and optical pumping with a 980 nm pump laser (dashed lines).  Optical component acronyms: fiber polarization controller (FPC), variable optical attenuator (VOA), single-mode fiber (SMF), $1300$ nm band reflected signal photodetector (PD$_{R-1300}$), $1300$ nm band transmitted signal photodetector (PD$_{T-1300}$), and $980$ nm band transmitted signal photodetector (PD$_{T-980}$).}
\label{fig:expt_configs}
\end{center}
\end{figure}

As described elsewhere \cite{ref:Srinivasan14}, the fiber taper
waveguide coupler (mounted as shown schematically in Fig.
\ref{fig:taper_cav_coupling}(a)) and microdisk chip are
incorporated into a customized continuous flow liquid He cryostat
that uses piezo-actuated slip-stick and flexure stages to achieve
precise cavity-taper alignment. The fiber pigtails are fed out of
the cryostat and are connected to a number of fiber optic elements
(Fig. \ref{fig:expt_configs}), such as fiber polarization-controlling
paddle wheels (FPC), variable optical attenuators (VOA), and fused fiber couplers,
which in total create an all-fiber setup where the only free-space
optics needed are for imaging. The setup has been configured so
that it is simple to switch between photoluminescence measurements
and resonant transmission and reflection measurements.

Photoluminescence (PL) is performed through use of a $980$ nm band
external cavity tunable diode laser as a pump source (Fig.
\ref{fig:expt_configs}(a)). The pump laser emission is directed
into the fiber taper after first going through an attenuator, a
50:50 coupler, and the 10 $\%$ port of a 10:90 fiber coupler. The
fiber taper is aligned along the microdisk edge, exciting
whispering gallery mode (WGM) resonances that are mapped out as a
function of wavelength by tuning the laser and recording the
transmitted signal with an InGaAs avalanche photodiode (APD). The
pump laser is fixed on-resonance with a WGM (which typically has a
$Q$ limited to $\approx10^3$ due to strong material absorption in
this wavelength band) and the resulting photoluminescence is
collected into both the forward and backward channels of the fiber
taper. The 90 $\%$ port of the 10:90 coupler is used to direct the
backward channel into a 550 mm grating spectrometer with a cooled
InGaAs linear array detector (512 elements, 25 $\mu$m pixel width)
and an effective spectral resolution of 35 pm at $\lambda=1.3$
$\mu$m.

Resonant transmission and reflection measurements (Fig.
\ref{fig:expt_configs}) are performed using a probe laser
consisting of an external cavity tunable diode laser (linewidth
$<$5 MHz) with continuous tuning in the $\lambda=1250$ nm to
$1330$ nm band. The probe laser beam is sent through an
attenuator, 50:50 coupler, and 10:90 coupler before entering the
fiber taper.  The transmitted signal is measured with a
thermoelectric-cooled InGaAs APD with a 1 kHz bandwidth, while the
reflected signal, previously directed to the grating spectrometer
in PL measurements, is measured with either a liquid nitrogen
cooled InGaAs APD (150 Hz bandwidth) or a second
thermoelectric-cooled InGaAs APD.  Although the experimental setup
allows for both the 980 nm and 1300 nm lasers to be simultaneously
directed into the microdisk cavity, for the experiments described
here, only one of these two lasers is used at a given time.  For
the pump-probe measurements described later (see Fig.
\ref{fig:bad_cavity_pump_probe_beam_saturation}(a)), a $1420$ to
$1500$ nm band external cavity tunable diode laser (linewidth $<$5
MHz) is used to excite the microdisk cavity while it is
simultaneously probed resonantly with the $1300$ nm laser beam.

The accuracy to which we can estimate intracavity photon number and
other important parameters of the coupled cavity-QD system using the
above experimental apparatus is dominated by uncertainties in the
optical loss within the various fiber optic components and the
calibration accuracy of optical power and frequency tuning of the
probe laser.  As given by eq. (\ref{eq:intracavity_photon_number_2}),
the intracavity photon number $n_{\text{cav}}$ depends upon
$P_{\text{in}}$, $\zeta$, ${\Delta}T$, and $Q_{T}$.  Uncertainty in
the input power \emph{at the microdisk} is related to the
sweep-to-sweep laser power fluctuations ($\pm 3$ $\%$), uncertainty in
the symmetry of the optical loss of the fiber taper about the
microdisk resonator ($\pm 5$ $\%$), and the variable optical loss in
the fiber unions between various fiber connections ($\pm 7.5$
$\%$). Knowledge of the actual dropped power into the resonator is
also affected by the uncertainty in ${\Delta}T$, which contains
contributions from noise in our detected signal ($\pm1$ $\%$) and the
degree to which the polarization of the input signal has been properly
aligned with the TE-polarized cavity mode ($\pm2.5$ $\%$). The
accuracy to which we know $Q_{T}$ is related to the accuracy of the
measured linewidth of the cavity mode, which is at the $\pm2$ $\%$
level due to our calibration error in the frequency tuning range of
the probe laser.  In sections \ref{sec:strong_coupling_data} and
\ref{sec:bad_cavity_data}, we plot quantities such as ${\Delta}T$ at a
fixed wavelength against $n_{\text{cav}}$, where the error bars
plotted for these quantities are calculated by propagating the above
uncertainties appropriately.

%The net result of an uncertainty analysis taking into account the
%values quoted here is that $n_{\text{cav}}$ and ${\Delta}T$ are
%known at a fractional $\pm10$ $\%$ to $\pm20$ $\%$ and $\pm5$ $\%$ to $\pm20$ $\%$ level,
%respectively.

\section{Device preparation}
\label{sec:device_preparation}

\subsection{Fabrication}

\begin{figure}[t]
\begin{center}
\includegraphics[width=\columnwidth]{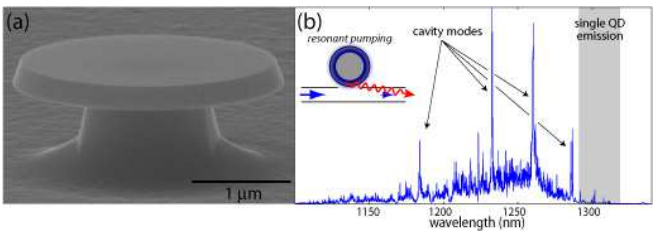}
\caption{(a) Scanning electron microscope image of a $D=2.5$
$\mu$m diameter microdisk cavity. (b) Photoluminescence spectrum
from a microdisk under free-space 830 nm excitation (150 nW
incident pump power) and fiber collection. Emission from the
excited and ground states of the QD ensemble is present from 1100
nm to 1310 nm, punctuated by regions of sharp enhanced emission
due to cavity modes. For $\lambda=1285$ nm - $1310$ nm, isolated
single QD emission is seen for a fraction of devices (15 $\%$). In
this device ($D=2$ $\mu$m), the cavity mode of interest is at
$\lambda=1287$ nm, $15$ nm blue detuned of an isolated QD exciton
state.}\label{fig:udisk_SEM_DWELL}
\end{center}
\end{figure}

The microdisks (Fig. \ref{fig:udisk_SEM_DWELL}(a)-(b)) are
fabricated in a material consisting of a single layer of InAs QDs
embedded in an In$_{0.15}$Ga$_{0.85}$As quantum well that resides
in a 256 nm thick GaAs waveguide.  This dot-in-a-well (DWELL)
epitaxy \cite{ref:Stintz,ref:Liu_G,ref:Cade1,ref:Cade2} is grown on top of a 1.5
$\mu$m thick Al$_{0.7}$Ga$_{0.3}$As layer that resides on a
semi-insulating GaAs substrate. In contrast to recent experiments
in which the QDs were grown with a low enough density that only a
single QD physically resided within the cavity
 \cite{ref:Badolato,ref:Hennessy3}, the density of QDs in the DWELL
material ( 300 $\mu$m$^{-2}$) means that there are on the order of
1000 QDs within a 2.5 $\mu$m diameter microdisk. To limit the
number of QDs that interact with the cavity mode of interest, we
fabricate microdisks of an appropriate diameter so that this mode
lies within the far red-detuned tail of the QD distribution (Fig.
\ref{fig:udisk_SEM_DWELL}), at a wavelength of $\approx1.3$ $\mu$m
where isolated single QD emission within this material was
previously observed \cite{ref:Srinivasan15}. A (top) disk diameter
$D=2.5$ $\mu$m is chosen in accordance with the results of finite
element method simulations \cite{ref:Srinivasan12} which indicate
that a TE$_{p=1,m=14}$ whispering gallery mode (WGM) is resonant
in the microdisk at $\lambda \approx 1300$ nm. The mode label
TE$_{p,m}$ corresponds to a mode of (dominantly) transverse
electric (TE) polarization, $p$ field antinodes in the radial
direction, and an azimuthal mode number $m$ corresponding to the
number of wavelengths around the circumference of the disk. The
TE$_{p=1,m=14}$ mode has a radiation-limited
$Q_{\text{rad}}>10^{8}$ (i.e., without taking into account
fabrication-induced roughness or material absorption) and an
effective standing wave mode volume of
$V_{\text{sw}}=3.2(\lambda/n)^3$ (disk refractive index $n=3.4$).
This effective mode volume can be used to estimate the strength of
coherent coupling between the exciton state of an embedded QD and
a cavity mode photon.  For a QD that is placed at a cavity field
antinode, whose exciton transition has an electric dipole parallel
to the cavity mode field polarization, and for a typical exciton
spontaneous emission lifetime $\tau_{\text{sp}}=$1 ns, the
coherent coupling rate is approximately $g_{0}/2\pi=$15 GHz.  This
value represents the \emph{maximum} expected coupling rate in this
system.  In comparison, a typical QD dephasing rate is
$\gamma_{\perp}/2\pi = 1$ GHz \cite{ref:Bayer} (if strictly
radiatively broadened, $\gamma_{\perp}/2\pi \approx 0.1$ GHz), and
the cavity decay rate for $Q_{T}=10^5$ is $\kappa_{T}/2\pi\approx
1$ GHz, so that strong coupling ($g>(\kappa_{T},\gamma_{\perp})$)
is potentially achievable in these devices.

\subsection{Device identification}

For a given sample, typically consisting of 50 microdisk cavities,
the procedure for investigating its potential for strong coupling
is as follows: (1) Optical spectroscopy with the fiber taper
waveguide is used to identify the spectral position and $Q$ of
cavity modes at room temperature.  In the wavelength range of
interest, the wavelength blue shift between room and low
temperature is 17 nm. (2) PL measurements through the fiber taper
are performed at 8 K to 15 K, identifying the spectral position of
QD states.  Transmission and reflection measurements through the
fiber taper versus wavelength are performed to confirm the
spectral position of the cavity modes. (3) If the cavity mode of
interest is $\lesssim4$ nm blue of the QD exciton state, it can be
tuned into resonance $\textit{in-situ}$ by introducing N$_{2}$ gas
into the cryostat \cite{ref:Srinivasan14,ref:Mosor}. If the cavity
mode is red of the QD exciton, the sample is removed from the
cryostat and blue-shifted through a digital etching process
\cite{ref:Hennessy2}, and the steps are repeated.

\subsubsection{Room temperature cavity mode spectroscopy}

\begin{figure}[t]
\begin{center}
\includegraphics[width=\columnwidth]{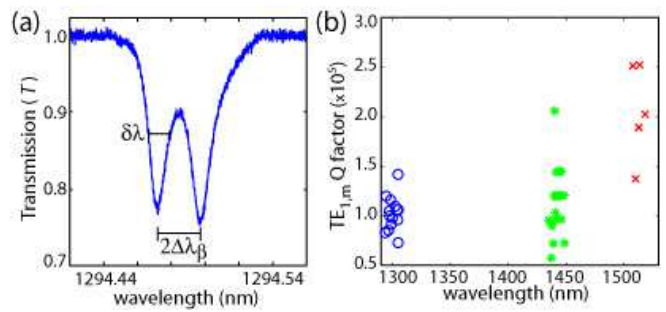}
\caption{(a) Normalized fiber taper transmission scan of a typical
TE$_{1,14}$ cavity mode of interest at a temperature T=15 K. The
pair of resonance dips is due to surface roughness which couples
and splits the clockwise and counterclockwise propagating modes of
the disk. (b) TE$_{1,m}$ cavity mode $Q$ for a number of $D=$2.5
$\mu$m disks in the 1300 nm band ($\textcolor{blue}{\circ}$) at
T=15 K, 1400 nm band ($\textcolor{green}{\ast}$) at T=298 K, and
1500 nm band ($\textcolor{red}{\times}$) at T=298
K.}\label{fig:cavity_spectroscopy}
\end{center}
\end{figure}

Room temperature cavity mode spectroscopy serves to eliminate from
consideration those devices whose cavity mode lies outside of the
wavelength range for which isolated single QD emission would
likely occur ($\lambda=1285$ nm -$1310$ nm).  It also gives an
indication of the cavity mode $Q$, although because the cavity
mode and the QD ensemble have different temperature-dependent
frequency shifts, the amount of absorption suffered by the cavity
at room temperature is different (and larger) than it is at low
temperature. Nevertheless, the room temperature measurement can at
least provide a lower bound on $Q$. For example, we have found
that, for a TE$_{1,14}$ mode at $\lambda=1300$ nm at cryogenic
temperatures, $Q=0.75\times10^5$ to $1.5\times10^5$ (Fig.
\ref{fig:cavity_spectroscopy}(a)-(b)), while at room temperature,
this mode is red-shifted by 17 nm and has $Q=0.6\times10^5$ to
$0.8\times10^5$.

We can gain further information about the cavity $Q$ by studying
TE$_{1,m}$ modes in a far red-detuned wavelength band
($\lambda=1420$ nm to $1565$ nm), so that we can largely eliminate
the effects of the QD absorption \footnote{This estimate is
possible because radiation losses for TE$_{1,m}$ remain
essentially negligible over this wavelength range for the
microdisk diameters we use, and because scattering losses remain
approximately constant.}. These modes differ in azimuthal mode
number by ${\Delta}m=2$ to ${\Delta}m=4$ in comparison to the
TE$_{1,14}$ mode at $\lambda\approx 1300$ nm. Using the fiber
taper to perform passive spectroscopy with the appropriate
external cavity tunable diode laser, we see (Fig.
\ref{fig:cavity_spectroscopy}(b)) that the $Q$ tends to increase
at longer wavelengths, with a highest $Q=2.5{\times}10^5$ at
$\lambda=1510$ nm ($V_{\text{eff}}=2.5(\lambda/n)^3$ at this
wavelength), approximately twice as large as the $Q$ in the 1300
nm band. These results indicate that the $Q$ of the mode of
interest at $\lambda=1300$ nm is absorption-limited, although
separating the loss into bulk and surface absorption components
requires further investigation.  In addition, these results are
consistent with those presented in Ref.
[\onlinecite{ref:Michael}], where absorption losses were seen to
increase as the wavelength was reduced from 1600 nm to 960 nm.

As seen in Fig. \ref{fig:cavity_spectroscopy}(a), even in absence
of coupling to a QD, the microdisk WGMs in our structures
typically appear as a resonance doublet.  This doublet structure
is due to surface roughness that couples and splits the initially
degenerate clockwise and counterclockwise propagating modes of the
disk, and has been observed in numerous WGM cavities
\cite{ref:Weiss,ref:Kippenberg,ref:Borselli2,ref:Srinivasan9}. To
some extent, the presence of this doublet splitting complicates
the nature of cavity-QD interactions within this system - both the
presence of the second cavity mode and passive modal coupling due
to the surface roughness must be taken into account to properly
describe the system.  This has been outlined in the quantum master
equation model developed in Ref. [\onlinecite{ref:Srinivasan13}], and is
re-considered qualitatively in Section \ref{subsec:modeling},
where we discuss numerical modeling of experimental results.

\subsubsection{Low temperature fiber-based photoluminescence measurements}

After room temperature cavity spectroscopy has been performed, the
sample is cooled down between 8 K to 15 K and those devices for
which cavity modes are appropriately spectrally positioned are
investigated in PL.  In comparison to more conventional free-space
PL measurements, the fiber-based method we use offers two
advantages \cite{ref:Srinivasan15}.  The first is an improvement
in collection efficiency by nearly 1 to 2 orders of magnitude over
our free-space measurements.  The second is the spatial
selectivity gained by pumping the microdisk through a 980 nm band
WGM.  By limiting the pumping area to the periphery of the
microdisk, we selectively excite those QDs which are most likely
to be overlapped with the 1300 nm cavity mode of interest.  Even
better spatial selectivity can be achieved, for example, by
pumping the cavity on a WGM that is separated in $m$-number by one
from the 1300 nm mode, so that the radial spatial overlap between
the two modes is nearly perfect.

\begin{figure}[t]
\begin{center}
\includegraphics[width=\columnwidth]{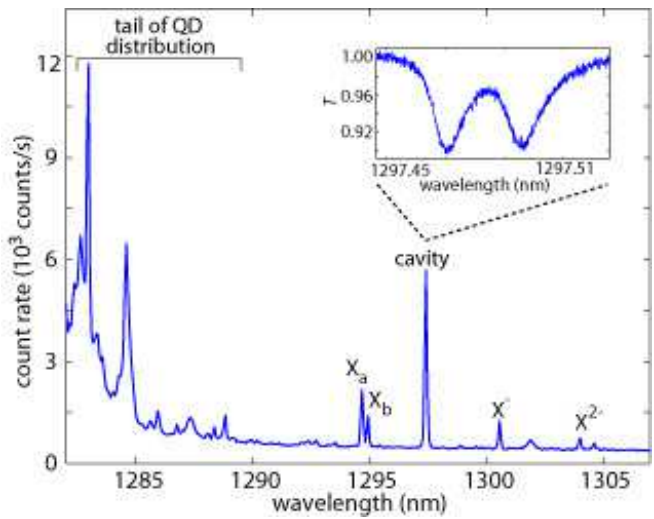}
\caption{Fiber-collected PL spectrum under pumping of a 980 nm
band WGM (5 nW input power into the fiber taper).  QD states are
labeled as: $X_{a}/X_{b}$ (fine-structure split neutral exciton
states), $X^{-}$ (negatively charged exciton), and $X^{2-}$
(double negatively charged exciton). The inset shows a normalized
transmission scan of the cavity mode at $\lambda=1297.5$ nm.
Sample temperature = 15 K.}\label{fig:PL_res_pumping}
\end{center}
\end{figure}

A PL spectrum for a device under resonant pumping of a 980 nm WGM
is shown in Fig. \ref{fig:PL_res_pumping}.  QD states are
identified in accordance with the procedure followed in Ref.
 [\onlinecite{ref:Srinivasan15}], which relied upon measurements of the
pump power dependence of the emission into each state and the
spectral splittings between the states, as well as the close
correspondence with previous spectroscopy of single DWELL QDs by
other researchers \cite{ref:Cade1,ref:Cade2}.  Of particular
interest to this work are the two neutral exciton lines of the QD,
$X_{a}$ and $X_{b}$, which correspond to orthogonally polarized
transitions split by the anisotropic electron-hole exchange
interaction resulting from asymmetries in the QD \cite{ref:Cade2}.
Cavity modes are typically identifiable in PL as well (emission
from cavity modes can be seen even when the cavity and QD are
far-detuned \cite{ref:Hennessy3,ref:Strauf2}, and may result from
the enhanced density of states for an ultra-small volume cavity
mode), and can be unambiguously confirmed through resonant
spectroscopy with the 1300 nm tunable laser.

Only a small fraction of devices (5 $\%$) produce a PL spectrum
like that in Fig. \ref{fig:PL_res_pumping}, where the cavity mode
is within a few nm of the isolated exciton lines of a single QD.
As discussed above, this low yield is due to the necessity of
working in the red-detuned tail of the QD distribution to limit
the number of QDs that are spectrally near the cavity mode of
interest and the random positioning of the cavity with respect to
these QDs.

\subsubsection{Cavity mode tuning}

We employ two mechanisms to tune the cavity modes.  Digital
etching \cite{ref:Hennessy2} outside of the cryostat is used to
blue-shift the cavity modes in discrete increments, while N$_{2}$
adsorption within the cryostat \cite{ref:Srinivasan14,ref:Mosor}
provides essentially continuous (red) tuning over a range of about
4 nm.  As the N$_{2}$ tuning method allows for real-time
monitoring of cavity-QD interactions, the desired cavity mode
position after fabrication (and any subsequent digital etching) is
within a few tenths of a nm blue of the QD exciton state, so that
the system can be effectively studied as a function of cavity-QD
detuning across resonance \footnote{The N$_{2}$ process produces a
wavelength shift as great as 4 nm, so that the cavity mode can be
further blue of the QD exciton and still be tuned into resonance.
However, at those tuning levels the cavity $Q$ is also
significantly degraded, by as much as a factor of 3 as studied in
Ref. [\onlinecite{ref:Srinivasan14}].}.

\begin{figure}[t]
\begin{center}
\includegraphics[width=\columnwidth]{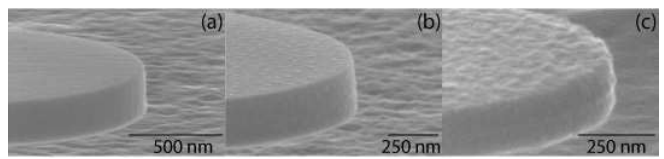}
\caption{SEM images of microdisk cavities (a) after initial device
fabrication, (b) after two steps of
H$_{2}$O$_{2}$/C$_{6}$H$_{8}$O$_{7}$ etching, (c) after fourteen
steps of H$_{2}$O$_{2}$/C$_{6}$H$_{8}$O$_{7}$
etching.}\label{fig:peroxide_citric_acid_etches}
\end{center}
\end{figure}

The digital etching process consists of alternating steps of
oxidation, either in atmosphere or in hydrogen peroxide
(H$_{2}$O$_{2}$), and oxide removal through a 1 molar solution of
citric acid (C$_{6}$H$_{8}$O$_{7}$). The native
oxidation/C$_{6}$H$_{8}$O$_{7}$ process produces a relatively
small cavity mode blue-shift of 0.8 nm per cycle, and does not
degrade the cavity $Q$ for the devices studied ($Q=10^5$) and the
number of etch cycles investigated (up to 6). The
H$_{2}$O$_{2}$/C$_{6}$H$_{8}$O$_{7}$ process produces a much
larger cavity mode blue-shift of 4.5 nm per cycle for a small
number of cycles ($\leq3$), and increases the cavity mode
linewidth by $<10$ $\%$. As the number of cycles increases
further, the amount of blue-shift per cycle increases ($\approx42$
nm for 6 cycles), as does the increase in cavity mode linewidth
($\approx25$ $\%$ for 6 cycles). SEM images of the microdisks
during this process (Fig. \ref{fig:peroxide_citric_acid_etches})
indicate that the degradation in $Q$ should not be too surprising
- even after 2 cycles of the H$_{2}$O$_{2}$/C$_{6}$H$_{8}$O$_{7}$,
the microdisk surface is noticeably altered (Fig.
\ref{fig:peroxide_citric_acid_etches}(b)), and after a large
number of cycles, the damage to the disk is quite significant
(Fig. \ref{fig:peroxide_citric_acid_etches}(c)).

\section{Coherent optical spectroscopy of a strongly coupled microdisk-QD system}
\label{sec:strong_coupling_data}

In this section, we present detailed measurements and analysis of
the system studied in Ref. [\onlinecite{ref:Srinivasan16}]. Although
some amount of repetition is necessary to provide background and
context, we have attempted to minimize this and will instead rely
upon citing the previous work when appropriate.

\subsection{Vacuum Rabi splitting measurements}
\label{subsec:Rabi_splitting_sc}

The device whose PL spectrum is shown in Fig.
\ref{fig:PL_res_pumping} presents a clear candidate for strong
coupling measurements.  Four cycles of the native
oxidation/C$_{6}$H$_{8}$O$_{7}$ process are used to blue-shift the
cavity mode, initially 3 nm red-detuned (see Fig.
\ref{fig:PL_res_pumping}), so that it is 260 pm blue detuned of
the QD $X_{a}$ neutral exciton line.  At this spectral position,
the cavity mode can easily be tuned into resonance with the
neutral ground-state QD exciton lines, and the transmitted and
reflected signals can be monitored by employing the experimental
setup depicted in Fig. \ref{fig:expt_configs}(a).

\begin{figure}
\begin{center}
\includegraphics[width=\columnwidth]{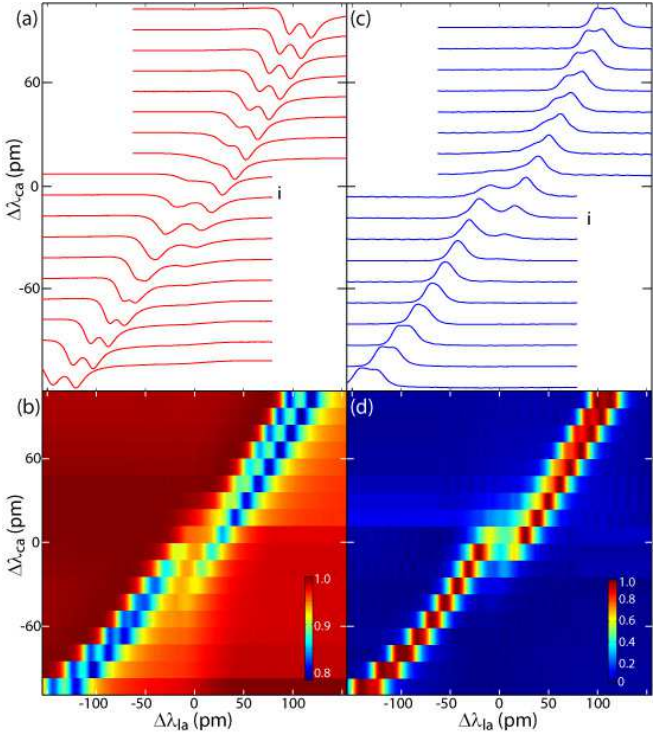}
\caption{(a-b) Transmitted and (c-d) reflected spectra from the
cavity as a function of laser-QD detuning
($\Delta\lambda_{\text{la}}$) and cavity-QD detuning
($\Delta\lambda_{\text{ca}}$), under weak driving by the probe
beam ($n_{\text{cav}}\ll1$) at 15 K.
 Spectra are normalized to unity.  The top plots (a,c) show a series of individual spectra,
 while the bottom plots (b,d) show a compilation of this data as an image plot.} \label{fig:trans_refl_spectra}
\end{center}
\end{figure}

Although continuous observation of the cavity response during the
$N_{2}$ adsorption process can be done \cite{ref:Srinivasan14},
and is limited only by the detector bandwidth (1 kHz) and data
acquisition rate, maintaining an adequate signal-to-noise ratio
requires us to average 10 to 20 single scans for every cavity
tuning point.  As a result, the cavity is tuned in discrete steps
by opening the valve between the N$_{2}$ chamber and the cryostat
for a fixed time (5 s), waiting until the cavity mode position
stabilizes, averaging the signal, and repeating. Over relatively
small tuning ranges, the amount of tuning per step is
approximately constant, and we typically choose a value of
(12$\pm$3) pm per step, which is controlled by adjusting the
N$_{2}$ flow rate.

Figure \ref{fig:trans_refl_spectra} presents a series of spectra
showing the normalized (to unity) transmitted and reflected signal
from the cavity over a tuning range of 240 pm, where the cavity is
driven by the tunable laser with an input power of 470 pW, so that
$n_{\text{cav}}\approx0.03$ and the system is well within the weak
driving limit. In this figure, the transmitted and reflected
signals are plotted against laser-QD detuning
($\Delta\lambda_{\text{la}}=\lambda_{\text{l}}-\lambda_{\text{a}}$)
and cavity-QD detuning
($\Delta\lambda_{\text{ca}}=\lambda_{\text{c}}-\lambda_{\text{a}}$).
As the pair of doublet cavity modes are tuned towards the short
wavelength neutral exciton line ($X_{a}$) of the QD, they undergo
a significant change in lineshape. Interaction with the $X_{a}$
line causes the long wavelength cavity mode of the doublet pair to
tune at a slower rate than the short wavelength cavity mode,
resulting in the formation of a singlet resonance. In addition, a
third resonance peak associated with the $X_{a}$ line begins to
appear on the red-side of the cavity modes. Further tuning of the
mode on to resonance with the QD results in the anti-crossing and
spectral splitting (vacuum Rabi splitting) that are most easily
visible in the image plots of Figure \ref{fig:trans_refl_spectra}
- in comparison to the bare-cavity (far-detuned from the QD) mode
spectrum where the doublet shape is due to surface roughness which
induces modal coupling between the propagating modes of the disk,
the doublet spectrum seen when the cavity and QD are resonant is
due to exciton-mode coupling, with a spectral splitting that is
much larger. The magnitude of this splitting relative to the peak
linewidths indicates that the system is in the strong coupling
regime ($g>(\kappa_{T},\gamma_{\perp})$), which is confirmed by
the detailed analysis of the data presented in section
\ref{subsec:modeling}.  Once the cavity is tuned sufficiently far
past the $X_{a}$ line, it regains its initial bare-cavity shape.

\begin{figure}
\begin{center}
\includegraphics[width=\columnwidth]{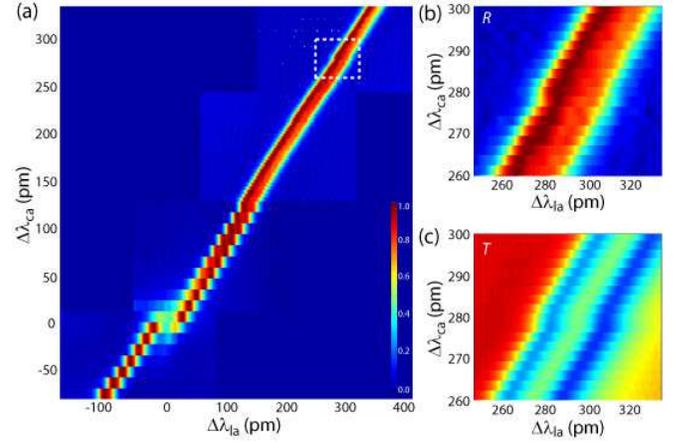}
\caption{(a) Normalized (to unity) reflected spectra from the
cavity as a function of laser-QD detuning
($\Delta\lambda_{\text{la}}$) and cavity-QD detuning
($\Delta\lambda_{\text{ca}}$). Normalized reflected (b) and
transmitted (c) spectra over a zoomed-in region (dashed box
region of (a)), where the cavity mode dispersively couples to the
$X_{b}$ state.} \label{fig:Xa_Xb_tuning}
\end{center}
\end{figure}

The N$_{2}$ tuning mechanism allows for the cavity to be easily
tuned through the other fine-structure split neutral exciton line
of the QD, the $X_{b}$ line (see Fig. \ref{fig:PL_res_pumping}).
A series of reflected spectra showing the cavity mode tuned
through both the $X_{a}$ and $X_{b}$ lines is shown in Fig.
\ref{fig:Xa_Xb_tuning}. In this data set, tuning through the
$X_{a}$ line was done with a relatively large step size, after
which the step size was reduced. Due to the extent of the tuning,
after the $X_{a}$ line is crossed, the step size is no longer
constant, but instead changes nonlinearly \cite{ref:Srinivasan14}.
A nonlinear (quadratic) transformation to convert between N$_{2}$
step number and cavity mode detuning was used to produce Fig.
\ref{fig:Xa_Xb_tuning} from the raw data.

%We first show (Fig. \ref{fig:Xa_Xb_tuning}(a)) the reflected
%spectra as a function of N$_{2}$ step number; in this set of data,
%tuning through $X_{a}$ was done with a relatively large step size,
%after which the step size was reduced.  Due to the extent of the
%tuning, after $X_{a}$ is crossed, the step size is no longer
%constant, but instead changes nonlinearly \cite{ref:Srinivasan14}.
%Figure \ref{fig:Xa_Xb_tuning}(b) plots the data as a function of
%cavity detuning, where a nonlinear (quadratic) transformation to
%convert between N$_{2}$ step number and cavity mode position was
%employed.

Tuning past the $X_{a}$ line, we see the cavity mode tune smoothly
and without interruption, until we observe a small frequency shift
that appears as a kink in the spectrum at
$\Delta\lambda_{\text{la}}\approx280$ pm past the $X_{a}$
transition. This spectrally corresponds to the position of the
$X_{b}$ line, with the small dispersive shift due to the cavity
weakly coupling to it.  The relative strength of coupling between
the cavity and the $X_{a}$ and $X_{b}$ lines can be related to the
polarization of the cavity mode, which is primarily oriented along
the radial direction of the microdisk.  The strong coupling of the
$X_{a}$ line to the cavity mode indicates that this transition is
polarized along the radial direction.  The $X_{b}$ transition is
orthogonally polarized to the $X_{a}$ transition, and thus is
coupled to the cavity mode through its weaker azimuthal electric
field component. As the polarization of the fine-structure split
neutral exciton lines is related to the geometry and orientation
of the QD within the InGaAs/GaAs host \cite{ref:Cade2}, the
relative coupling strengths of the $X_{a}$ and $X_{b}$ exciton
lines is an indication of the QD orientation along the radial
direction of the microdisk.

\subsection{Nonlinear spectroscopy}
\label{subsec:nonlinear_sc}

The physical trapping of the QD within a solid material means that
complete spectral characterization of the cavity-QD system as a
function of laser power can be made.  In comparison, such a
measurement is much more challenging in atomic physics based
experiments, where measurements on \emph{one-and-the-same} atom
are limited by trap times of 1 s to 10 s \cite{ref:Boca}.
Reference [\onlinecite{ref:Srinivasan16}] presented measurements
of the reflected spectrum near resonance
($\Delta\lambda_{\text{ca}}\approx-$12 pm), and showed that
saturation occurs for $n_{\text{cav}}\gtrsim0.1$. Here, we
complement that work with measurements of the transmitted
spectrum, and show how the transmission contrast can be varied
significantly by adding less than one intracavity photon to the
system.

\begin{figure}
\begin{center}
\includegraphics[width=\columnwidth]{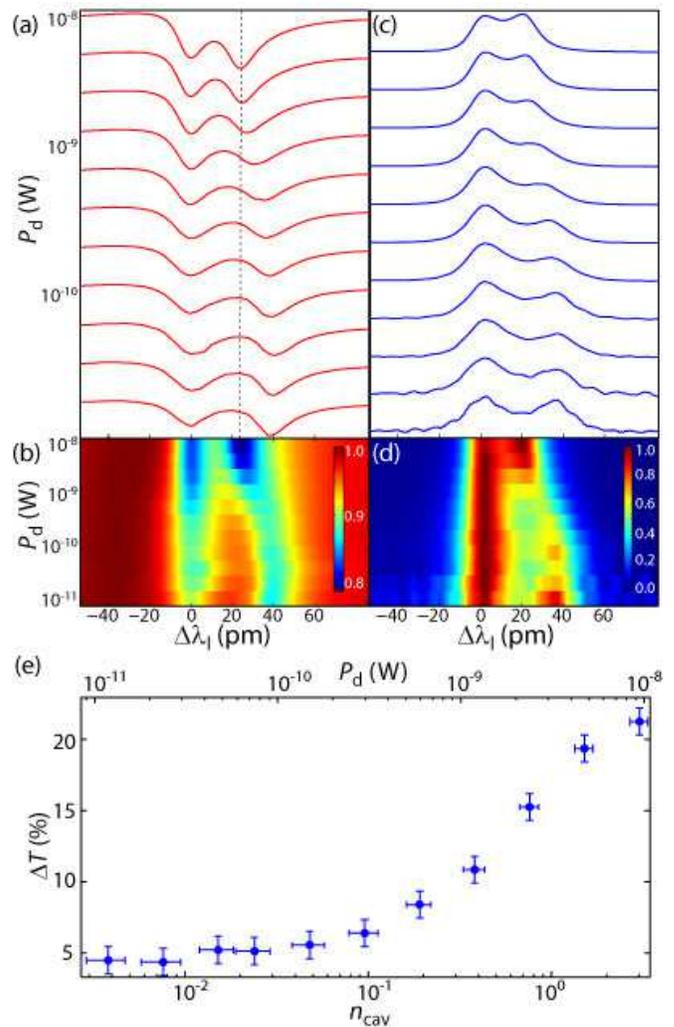}
\caption{Normalized (to unity) transmitted (a-b) and reflected (c-d)
spectra from the cavity as a function of dropped power into the
cavity ($P_{\text{d}}$) at the cavity-QD detuning level marked 'i'
in Fig. \ref{fig:trans_refl_spectra}(a),(c). Here, the spectra are
plotted as a function of $\Delta\lambda_{l}$, the detuning from
the short-wavelength resonance peak/dip. (e) Normalized
transmission contrast (${\Delta}T$) vs. $P_{\text{d}}$ and average
intracavity photon number ($n_{\text{cav}}$), taken at the
dashed-line spectral position shown in part (a).}
\label{fig:power_series_data}
\end{center}
\end{figure}

We first re-consider the QD-cavity system close to resonance,
labeled as scan 'i' in Fig. \ref{fig:trans_refl_spectra}(a),(c).
Varying the dropped power $P_{\text{d}}$ into the cavity between
10 pW and 10 nW, we plot a series of transmitted and reflected
spectra in Fig. \ref{fig:power_series_data}(a)-(d).  Here, the
spectra are plotted as a function of laser detuning from the short
wavelength peak of the doublet.  As $P_{\text{d}}$ increases, the
spectral splitting decreases and the spectra saturate towards
their bare-cavity shape.  This saturation can perhaps best be seen
by plotting a specific spectral feature against $P_{\text{d}}$ and
$n_{\text{cav}}$, determined through eq.
(\ref{eq:intracavity_photon_number}). In Ref.
[\onlinecite{ref:Srinivasan16}], the spectral splitting in the
reflected signal and the peak reflected signal level were
examined.  We supplement that here in Fig.
\ref{fig:power_series_data}(e) by following the transmission
contrast (${\Delta}T$) at the bare-cavity position of the longer
wavelength doublet resonance (marked by a dashed line in Fig.
\ref{fig:power_series_data}(a)).  The data shows saturation for
$n_{\text{cav}}\gtrsim0.1$, with ${\Delta}T$ increasing by over a
factor of $4$ when going from weak to strong driving.  Larger
changes in the relative transmission contrast between the weak and
strong driving regimes can be achieved by adjusting the cavity-QD
detuning level and judiciously choosing the wavelength at which
the transmission contrast is recorded.

For low power switching applications, the system studied here has
a number of potentially important advantages.  Along with the
strong optical nonlinearity available at the single photon level
and the robustness of using a monolithic, semiconductor system,
the fiber coupling method translates the small intracavity photon
numbers at which saturation occurs into low input powers. This is
a key point, as it is the parameters of the QD-cavity system that
set the intracavity photon number at which saturation occurs, but
it is the efficiency of coupling into the cavity (quantified by a
coupling parameter $K$ given by $\kappa_{e}/\kappa_{i+P}$, the
ratio of waveguide-cavity coupling to the total of intrinsic and
parasitic loss) that determines the input power at which this
intracavity photon number occurs. In the devices studied here, the
input power into the taper $P_{\text{in}}$ is only about a factor
of 5 larger than $P_{\text{d}}$, so that saturation occurs for
$P_{\text{in}}<1$ nW.

Clearly, investigation of the time-dependent characteristics of
this system is necessary before too much more can be said about
switching applications.  Nevertheless, a few comments can be made
on the basis of the steady-state nonlinear spectroscopy we have
performed.  Improving the ratio of ${\Delta}T$ in the strong and
weak driving regimes will be important in ultimately being able to
discriminate between the 'on' and 'off' response of the system.
The key ingredient in such an optimization would be an improvement
in the taper-cavity coupling, which can straightforwardly be
achieved through slightly smaller diameter cavities.  In addition,
the reflected signal may ultimately be a preferred option, due to
the comparative ease with which a null signal can be generated.
However, effective use of the reflected signal will likely require
larger absolute reflection values; as discussed in the following
section, the peak reflected signal (normalized to input power) for
the device discussed here is 0.6 $\%$.

%A larger reflected signal could be generated by improving the
%system parameters (e.g taper-cavity coupling, cavity intrinsic
%loss rate, QD dephasing rate) and in particular, the amount of
%clockwise-counterclockwise mode coupling.

\subsection{Numerical modeling}
\label{subsec:modeling}

\renewcommand{\arraystretch}{1.3}
\renewcommand{\extrarowheight}{0pt}
\begin{table}[t]
\caption{Parameters in the quantum master equation model.}
\label{table:cQED_quantities}
%\begin{center}
%\begin{tabular}{|c|c|c|c|c|c|c|c|c|}
\begin{tabularx}{\linewidth}{lX}
\hline \hline
Symbol & Description \\
%\hhline{|=:=:|}
\hline
$\text{P}_{i/R/T}$  & incident/reflected/transmitted signal \\
$\kappa_{e/i}$      & cavity decay rate due to waveguide
coupling/intrinsic loss \\
$\kappa_{T}$        & total cavity loss rate = 2$\kappa_{e}$ +
$\kappa_{i}$ \\
$\text{a}_{\text{cw/ccw}}$ & cavity clockwise/counterclockwise
field amplitude \\
$g_{\text{tw}}$ & cavity-QD coupling rate for traveling wave modes \\
$\gamma_{\parallel}$ & QD energy dephasing rate \\
$\gamma_{\text{p}}$  & QD non-radiative decay rate\\
$\gamma_{\perp}$     & QD transverse decay rate =
$\gamma_{\parallel}$/2 + $\gamma_{\text{p}}$ \\
$\gamma_{\beta}$     & surface roughness induced cw/ccw mode coupling\\
$\xi$                & relative phase between surface scattering
and exciton-mode coupling \\
$g_{\text{sw1,2}}$ & cavity-QD coupling rate for standing wave modes, known from $g_{\text{tw}}$ and $\xi$\\

\end{tabularx}
%\end{center}
\end{table}
\renewcommand{\arraystretch}{1.0}
\renewcommand{\extrarowheight}{0pt}

\begin{figure}[t]
\begin{center}
\includegraphics[width=0.625\columnwidth]{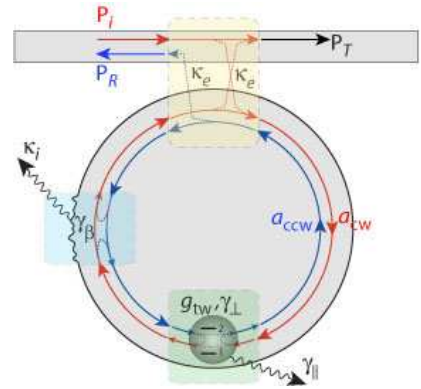}
\caption{Illustration of the processes considered in modeling the
microdisk-QD system.} \label{fig:cQED_schematic}
\end{center}
\end{figure}

The steady-state behavior of the system is modeled using a quantum
master equation (QME) approach, as described in detail in Ref.
[\onlinecite{ref:Srinivasan13}].  The standard picture of single
mode cavity QED \cite{ref:Carmichael} with an atom-cavity coupling
strength $g$, cavity loss rate $\kappa$, and atomic decay via
spontaneous emission (spontaneous emission lifetime
$\tau_{\text{sp}}$) is augmented to include a number of features
specific to the system we study. These include: (1) a second
cavity mode, as WGM microcavities support degenerate clockwise
($cw$) and counterclockwise ($ccw$) propagating modes; (2) modal
coupling at a rate $\gamma_{\beta}$ between the $cw$ and $ccw$
modes, due to fabrication-induced surface roughness; (3)
additional decay channels for the QD, where along with radiative
decay at a rate $1/\tau_{\text{sp}}$ we consider additional
inelastic dephasing processes so that the QD energy dephasing rate
$\gamma_{\parallel}$ can be larger than $1/\tau_{\text{sp}}$. We
also consider the possibility of pure elastic dephasing at a rate
$\gamma_{p}$, with the total transverse dephasing of the QD being
$\gamma_{\perp}=\gamma_{\parallel}/2+\gamma_{p}$; and (4)
input-output coupling to the optical fiber taper waveguide, so
that the total cavity decay rate $\kappa_{T}$ is split into an
intrinsic loss component $\kappa_{i}$ and a coupling decay rate
$\kappa_{e}$ into each of the forward and backward channels of the
fundamental mode of the fiber taper.  Figure
\ref{fig:cQED_schematic} schematically describes the system
considered in the model, while Table \ref{table:cQED_quantities}
lists the relevant parameters involved.

\begin{figure*}
\begin{center}
\includegraphics[width=1.75\columnwidth]{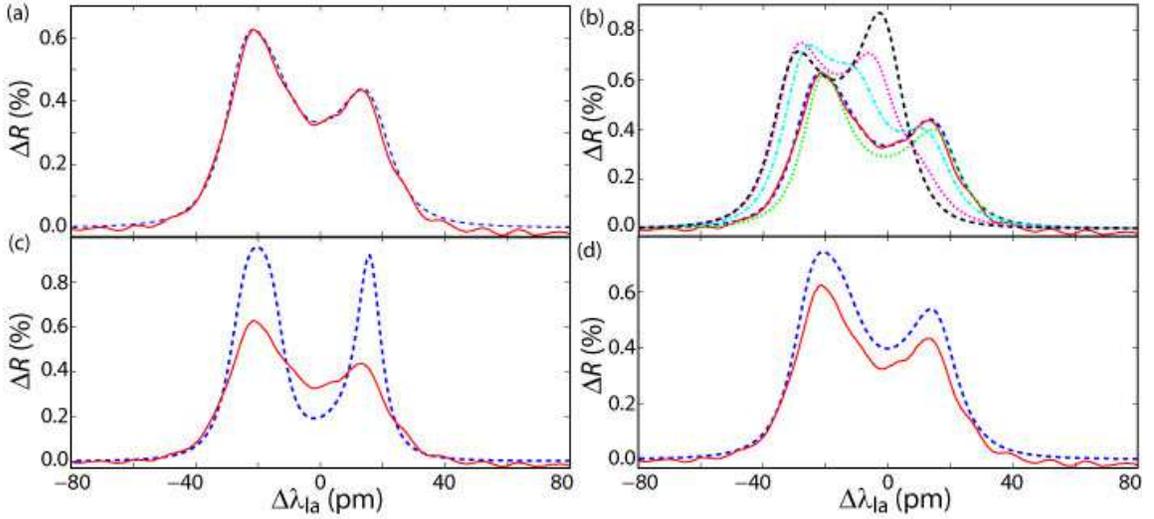}
\caption{(a) Reflection spectrum from the QD-microdisk system near
resonance (position (i) in Fig.
  \ref{fig:trans_refl_spectra}(c)) under weak driving.  The solid red
  line is the measured reflected power normalized to input power; the
  dashed blue line is a QME model of the system using the parameters listed in Table
\ref{table:cQED_fit_params}. (b)-(d) Experimental data (red solid
line) and model plots (dashed lines) for variations to the
parameters listed in Table \ref{table:cQED_fit_params} as follows:
(b) $\xi=\{0,0.25,0.5,0.75,1\}\pi$ corresponding to dotted green,
dashed blue, dash-dotted cyan, dotted purple, and dashed black
lines, respectively, (c)
$\gamma_{\perp}/2\pi=(\gamma_{\parallel}/2)/2\pi=(1/2\tau_{\text{sp}})/2\pi$=0.08
GHz, and (d) $\gamma_{\perp}/2\pi=(\gamma_{\parallel}/2
+\gamma_{p})/2\pi=(1/2\tau_{\text{sp}}+\gamma_{p})/2\pi$=0.97
GHz.}\label{fig:refl_signal_expt_vs_theory}
\end{center}
\end{figure*}

Figure \ref{fig:refl_signal_expt_vs_theory}(a) compares simulation
and experimental results for the scan marked 'i' in Fig.
\ref{fig:trans_refl_spectra}(c), with the values of the relevant
physical parameters used in the model listed in Table
\ref{table:cQED_fit_params}.  Importantly, the comparison of the
simulated and experimental results is done on the absolute
reflected signal (${\Delta}R$), which is normalized to input power
rather than unity, so that both the shape and amplitude of the
signal are considered in fitting the data. Examining the reflected
spectrum has benefit in that it is a direct probe of the
intracavity field (suppressing all non-resonant light not coupled
into the cavity-QD system).  In Table \ref{table:cQED_fit_params}
we have noted the 'source' of the parameter, whether it be from
the bare-cavity transmission spectrum or if it is obtained through
a fit to the measured resonant cavity-QD reflection and
transmission data. The listed uncertainty value for each parameter
is that due to the uncertainty in the measured data (such as
wavelength detuning and/or reflected signal amplitude) and the
corresponding range of parameter values that fit the data to
within this uncertainty range.

The bare-cavity spectrum, far blue-detuned of the QD exciton, can be
fit with a simple coupled mode theory (CMT) model.  The CMT model
gives directly the total cavity decay rate $\kappa_{T}$ and
waveguide-cavity coupling depth ${\Delta}T$, from which $\kappa_{i}$
and $\kappa_{e}$ are determined.  The CMT model also yields the
amplitude of the modal coupling rate $|\gamma_{\beta}|$, corresponding
to the doublet mode splitting seen in Fig.
\ref{fig:cavity_spectroscopy}(a).

The rest of the parameters, associated with cavity mode coupling
to the excitonic state of the QD, are determined by examining the
resonant reflection and transmission data, and fitting it to the
QME model.  The cavity-QD detuning $\Delta\lambda_{\text{ca}}$ is
known by determining the position of resonance, through an
anti-crossing measurement such as in Fig.
\ref{fig:trans_refl_spectra}, and knowledge of the approximate
wavelength tuning per N$_{2}$ adsorption step.  The relative
backscattering phase, $\xi$, is determined by the QME model fit to
the reflected and transmitted signal across the entire
anti-crossing curve of Fig. \ref{fig:trans_refl_spectra}.  Its
value is estimated, based upon the quality of the QME model fit,
to be $\xi=(0.25 \pm 0.05)\pi$.

The splitting between the branches of the anti-crossing diagram at
zero detuning is used to help determine the cavity-QD coupling
rate $g$.  Modal coupling between the $cw$ and $ccw$ traveling
wave modes is mediated by both surface roughness and the resonant
interaction with the exciton states of the QD, and as a result,
the cavity-exciton coupling rate for standing wave modes
$g_{\text{sw1,2}}$ depends on the relative phase between the two
($\xi$).  In particular, $g_{\text{sw1,2}}$ is modified by a
factor of $(1{\pm}e^{i\xi})/\sqrt{2}$ in comparison to the
coherent coupling to a purely travelling wave WGM
($g_{\text{tw}}$) \cite{ref:Srinivasan13}. For $\xi=0.25\pi$, the
exciton primarily couples to the longer wavelength standing wave
cavity mode, and the on-resonance vacuum Rabi-splitting therefore
very closely corresponds to $2g_{\text{sw,1}}$. We then determine
$g_{\text{sw,2}}$ and $g_{\text{tw}}$ from the above formula in
terms of $\xi$ and $g_{\text{sw,1}}$.

Finally, the decoherence of the excitonic state of the QD, as
determined by $\gamma_{p}$ and $\gamma_{\parallel}$, affect the width,
contrast, and amplitude of the Rabi-split peaks in the reflected
spectrum, providing us with a method for estimating each of their
values from the QME model.  Specifically, for the system under study,
the pure dephasing of the excitonic line ($\gamma_{p}$) sensitively
affects the linewidth and contrast of the reflected resonance peaks
and only weakly affects the overall amplitude of the reflected signal.
The population decay rate, $\gamma_{\parallel}$, in contrast, has a
significant impact on the resonant reflected signal amplitude through
additional loss added to the system, while contributing only weakly to
the resonance peak linewidths due to the relatively fast bare-cavity
decay rate.

%To determine the values for these parameters, we compute a series
%of spectra as a function of $\Delta\lambda_{\text{ca}}$ and
%compare the results with a series of experimental spectra, such as
%those in Fig. \ref{fig:trans_refl_spectra}.

\renewcommand{\arraystretch}{1.3}
\renewcommand{\extrarowheight}{0pt}
\begin{table*}
\caption{Values used in the QME fit in Figure \ref{fig:refl_signal_expt_vs_theory}.}
\label{table:cQED_fit_params}
\begin{center}
%\begin{tabular}{|c|c|c|c|c|c|c|c|c|}
\begin{tabularx}{0.55\linewidth}{lll}
\hline \hline
Parameter & \quad Value & \quad Source \\
%\hhline{|=:=:|}
\hline
$\kappa_{T}/2\pi$      & \quad $1.62 \pm 0.032$ GHz & \quad bare-cavity transmission \\
$\kappa_{e}/2\pi$      & \quad $0.17 \pm 0.0085$ GHz & \quad bare-cavity transmission \\
$\kappa_{i}/2\pi$      & \quad $1.27 \pm 0.064$ GHz & \quad bare-cavity transmission \\
$\gamma_{\beta}/2\pi$  & \quad $1.99 \pm 0.040$ GHz & \quad bare-cavity transmission \\
$\xi$                  & \quad $(0.25 \pm 0.05)\pi$ & \quad coupled cavity-QD tran./refl. spectrum \\
$g_{\text{sw,1}}/2\pi$ & \quad $2.93 \pm 0.15$ GHz & \quad coupled cavity-QD tran./refl. spectrum \\
$g_{\text{sw,2}}/2\pi$ & \quad $1.21 \pm 0.22$ GHz & \quad coupled cavity-QD tran./refl. spectrum \\
$g_{\text{tw}}/2\pi$   & \quad $2.24 \pm 0.20$ GHz & \quad coupled cavity-QD tran./refl. spectrum \\
$\gamma_{\parallel}/2\pi$ & \quad $0.55 \pm 0.28$ GHz & \quad coupled cavity-QD tran./refl. spectrum \\
$\gamma_{p}/2\pi$  & \quad $0.89 \pm 0.089$ GHz & \quad coupled cavity-QD tran./refl. spectrum \\
$\gamma_{\perp}/2\pi$  & \quad $1.17 \pm 0.32$ GHz & \quad coupled cavity-QD tran./refl. spectrum \\
\end{tabularx}
\end{center}
\end{table*}
\renewcommand{\arraystretch}{1.0}
\renewcommand{\extrarowheight}{0pt}

The results of our modeling confirm that the $X_{a}$ exciton line
and the TE$_{1,14}$ cavity mode are in the strong coupling regime,
with $g_{\text{sw,1}}/2\pi=$2.93 GHz, more than a factor of two
times larger than the average of $\kappa_{T}/2\pi=$1.62 GHz and
$\gamma_{\perp}/2\pi=$1.17 GHz \footnote{We have corrected two
typos
  in Ref. [\cite{ref:Srinivasan16}], where $\kappa_{T}$ and $\kappa_{i}$
  were incorrectly quoted.  The values that were quoted in that work
  were appropriate when the cavity was far-detuned (a few tenths of a
  nm) from the QD.  The corrected values are appropriate for the
  cavity mode just blue-detuned of the QD, and are larger due to
  additional loss induced by the N$_{2}$ tuning
  process \cite{ref:Srinivasan14}.}.  This can be compared to the
expected coherent coupling rate between cavity mode and exciton
line, given for a traveling wave mode by $g_{\text{tw}} =
\eta\sqrt{3c\lambda_{0}^2/8\pi{n^3}\tau_{\text{sp}}V_{\text{tw}}}$,
and as mentioned above, multiplied by a factor of
$(1{\pm}e^{i\xi})/\sqrt{2}$ for standing wave modes.  In this
expression, $\tau_{\text{sp}}$ is the spontaneous emission
lifetime of the excitonic transition, $c$ is the speed of light,
$n$ is the refractive index of the semiconductor host,
$\lambda_{0}$ is the emission wavelength, and $\eta$ takes into
account the position and orientation of the exciton dipole
($\eta=$1 in the case of ideal coupling; in practice, $\eta<1$).
$V_{\text{tw}}$ is the effective mode volume for a traveling wave
mode, and is 6.4$(\lambda/n)^3$ for the mode of interest in the
cavity under study, as determined by finite element method
simulations.  For a spontaneous emission lifetime of $\tau_{sp}=1$
ns, typical of similar QD excitons \cite{ref:Bayer}, the measured
value of $g_{\text{sw,1}}$ for coupling to the $X_{a}$ exciton
line is a little more than five times smaller than the estimated
maximum achievable value $g_{0}/2\pi=15$ GHz, so that
$\eta\approx0.2$. The measurements of weak coupling to the other
fine-structure split exciton line ($X_{b}$) in Section
\ref{subsec:Rabi_splitting_sc} indicate that the $X_{a}$ line is
relatively well-aligned in polarization with the predominantly
radially polarized cavity mode. As a result, we estimate that
spatial misalignment is the primary cause for the reduced value of
$g_{\text{sw,1}}$, with the QD approximately $300$ nm inwards from
the position of peak field strength (the small value of
$\xi=0.25\pi$ reduces $g_{\text{sw,1}}$ by $<10$ $\%$).

To illustrate the sensitivity of the model to the inferred
parameters, in Figure \ref{fig:refl_signal_expt_vs_theory}(b)-(d)
we present results in which the values for $\xi$,
$\gamma_{\perp}$, and $\gamma_{\parallel}$ are varied. In Fig.
\ref{fig:refl_signal_expt_vs_theory}(b), we plot the model results
for a wide range of values of $\xi$.  For these values, the
predicted reflected spectrum qualitatively changes.  In Fig.
\ref{fig:refl_signal_expt_vs_theory}(c), we show the calculated
reflected spectrum for the same set of parameters as in Table
\ref{table:cQED_quantities}, but under the assumption that the
only QD dephasing is due to spontaneous emission with a radiative
lifetime $\tau_{\text{sp}}$=1 ns, so that
$\gamma_{\perp}=\gamma_{\parallel}/2=1/2\tau_{\text{sp}}$.  We see
that the generated fit differs significantly from the data in
terms of the peak height, contrast, and width.  Figure
\ref{fig:refl_signal_expt_vs_theory}(d) re-instates the
phase-destroying collisional rate $\gamma_{p}$, but still assumes
all energy decay is due to radiative decay, so that
$\gamma_{\perp}=1/2\tau_{\text{sp}}+\gamma_{p}$.  Now, the model
fit is much closer to the data, particularly in terms of peak
contrast and width, but the absolute amplitude is too large.
Matching the absolute amplitude requires $\gamma_{\parallel}$ to
be increased, with $\gamma_{\parallel}/2\pi$=0.55 GHz producing
the model fit in Fig. \ref{fig:refl_signal_expt_vs_theory}(a)
which best matches the data. We have suggested that this increased
$\gamma_{\parallel}$ is purely due to non-radiative energy decay.
However, it is also possible that $\tau_{sp}$ could be shorter
than $1$ ns, which would then require the overlap between the
cavity and QD ($\eta$) to be smaller than we have estimated.
Future studies to better address the breakdown of radiative and
non-radiative components of $\gamma_{\parallel}$ could include
lifetime measurements of the QD when it is non-resonant with the
cavity and high spatial resolution measurements of the QD
position.

\begin{figure}
\begin{center}
\includegraphics[width=\columnwidth]{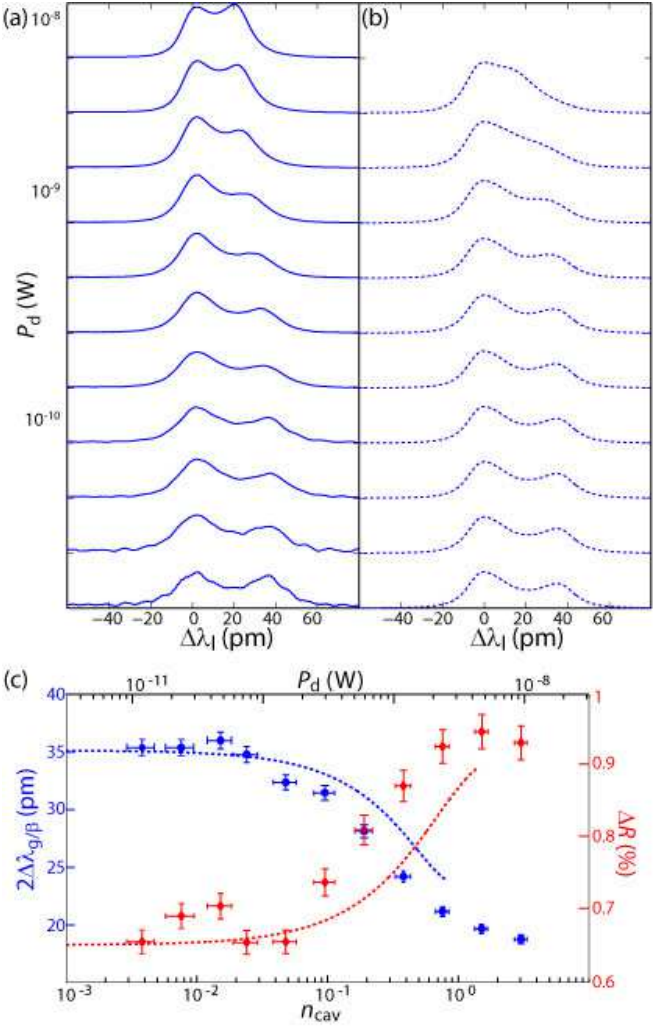}
\caption{(a) Experimental data and (b) theoretical model of the
normalized (to input power) reflected spectrum as a function of
dropped power into the cavity ($P_{\text{d}}$) at the cavity-QD
detuning level marked 'i' in Fig. \ref{fig:trans_refl_spectra}(c).
Here, the spectra are plotted as a function of
$\Delta\lambda_{l}$, the detuning from the short-wavelength
resonance peak/dip. (c) Mode splitting (circles) and peak
reflection value (diamonds) as a fuction of $P_{\text{d}}$ and
$n_{\text{cav}}$.  Theoretical predictions are shown as dashed
lines.} \label{fig:power_series_expt_vs_theory}
\end{center}
\end{figure}

Using the model parameters listed in Table
\ref{table:cQED_fit_params}, we next examine the nonlinear
behavior of the cavity reflection spectrum as a function of drive
strength in Fig. \ref{fig:power_series_expt_vs_theory}.  Just as
was done in Fig. \ref{fig:power_series_data}(c-d), in Fig.
\ref{fig:power_series_expt_vs_theory}(a) we plot reflection
spectra for a series of drive strengths (dropped power into the
cavity $P_{\text{d}}$), but here the spectra are normalized to
input power rather than to unity, thus allowing for a comparison of the recovery of the reflected signal \emph{amplitude}.  In Fig.
\ref{fig:power_series_expt_vs_theory}(b), we plot the reflection
spectra generated by the model for the same values of
$P_{\text{d}}$, where computer memory limitations have prevented us from
being able to accurately model the cavity mode at the highest
value of $P_{\text{d}}$ (due to the limited photon Fock space size of $8$ intra-cavity photons).  The model reproduces the features of the
data relatively well in terms of the peak reflection level
(${\Delta}R$), which steadily increases with larger values of
$P_{\text{d}}$, and the doublet splitting, which decreases from a
value dominated by exciton-mode coupling (2$\Delta\lambda_{g}$) to
one dominated by surface-roughness-induced backscattering
(2$\Delta\lambda_{\beta}$).  This behavior is quantified by
plotting the splitting and ${\Delta}R$ as a function of
$P_{\text{d}}$ and $n_{\text{cav}}$ in Fig.
\ref{fig:power_series_expt_vs_theory}(c). As was the case when
considering the transmission contrast in Fig.
\ref{fig:power_series_data}(e), we see the system begin to
saturate for $n_{\text{cav}}\approx0.1$.  The behavior of the
experimental data is qualitatively reproduced by the model,
although there are some discrepancies, such as the model
prediction of saturation occurring at somewhat higher power
levels. As discussed later in the paper, some portion of this
discrepancy may be due to the presence of an additional saturation
mechanism within these semiconductor-based devices.

Finally, a comment is warranted here with regard to the possible
effects of the limited time resolution of the measurements we have
conducted thus far and the accurate determination of $\gamma_{p}$
and $\gamma_{\parallel}$. The transmission and reflection spectra
data used in fitting to the QME model described above consist of
averages of 10 to 20 single wavelength scans.  In addition, the
photodetector signal representing transmission or reflection is
amplified with a low-pass pre-filter of bandwidth 30 Hz to 300 Hz
(depending upon the photodetector used) to remove electronic noise
from the detector pre-amplifier.  In order to faithfully represent
the optical response of the cavity-QD system, each wavelength scan
over a $600$ pm wavelength range is taken with a scanning
repetition rate of 0.3 Hz to 1 Hz (a piezo-stack is used to scan
the grating feedback wavelength of the external cavity laser
diode) and a sampling resolution of $0.5$ pm, corresponding to an
analog-to-digital sampling time of $1$ ms to $3$ ms for the data
acquisition card used to record the spectrum.  The time resolution
for our measurements is then seen to be limited by the amplifier
pre-filter bandwidth to a value on the order of 10 ms.  As a
result, it is conceivable that the measured peak height, width,
and contrast of the strongly coupled resonances of the QD-cavity
system are not entirely due to fast dephasing mechanisms as we
have assumed above in our QME model through the introduction of
$\gamma_{p}$ and $\gamma_{\parallel}$, but could instead be due to
slower processes such as the charging and discharging of the QD
(and its nearby environment) on the timescale of hundreds of ns to
ms \cite{ref:Santori4}.  The resulting spectral diffusion of the
QD exciton lines due to charge fluctuations would cause the time
averaged signal of a sharper, higher contrast set of resonance
peaks of the coupled system to smear out due to small shifts in
the exciton line or to be averaged with the bare-cavity response
due to larger shifts in the exciton line.  In the next section we
show both single scan and averaged data for a QD exciton line on
resonance with the cavity mode (see Fig.
\ref{fig:single_vs_avg_scans}), where the single scan data shows
evidence of fluctuations in the resonant transmission dips that
are significantly absent for that of the bare-cavity mode.  This
data suggests that our quantum master equation model does not
fully capture the behavior of our system, and processes such as
spectral diffusion of the exciton lines of the QD may be of
importance.

%$\approx290$ ps).  However, this would in turn mean that the peak
%achievable coupling strength would also be larger (by about a
%factor of $\sqrt{3}$), meaning that the

\section{Linear and nonlinear spectroscopy of a coupled microdisk-QD system in the bad cavity limit}
\label{sec:bad_cavity_data}

In this section, we present measurements of a coupled microdisk-QD
system in the bad cavity limit, where $\kappa_{T}$ exceeds $g$ and
$\gamma_{\perp}$.  As we shall see, $g$ is still large enough for
the system to exhibit many of the same properties we saw in the
strongly coupled system, such as anti-crossing and saturation for
$n_{\text{cav}}<1$. Along with serving as an additional test of
the experimental setup developed, the main motivation behind
presenting this data lies in section \ref{subsec:pump_probe},
where the system is studied in a pump-probe configuration, with an
eye towards potential future experiments involving control of the
QD through the ac-Stark effect and spectroscopy of higher states
of the Jaynes-Cummings system. In the measurements presented here,
a laser beam is coupled into a far red-detuned mode of the cavity
while the system is simultaneously probed near-resonance as it was
in section \ref{sec:strong_coupling_data}.  We observe a
saturation of the transmission/reflection spectrum with increasing
pump power, reminiscent of the saturation observed in section
\ref{subsec:nonlinear_sc} when increasing the probe power, but
occurring at intracavity photon numbers that are nearly one order
of magnitude larger.

\subsection{Vacuum Rabi splitting measurements}

\begin{figure}
\begin{center}
\includegraphics[width=\columnwidth]{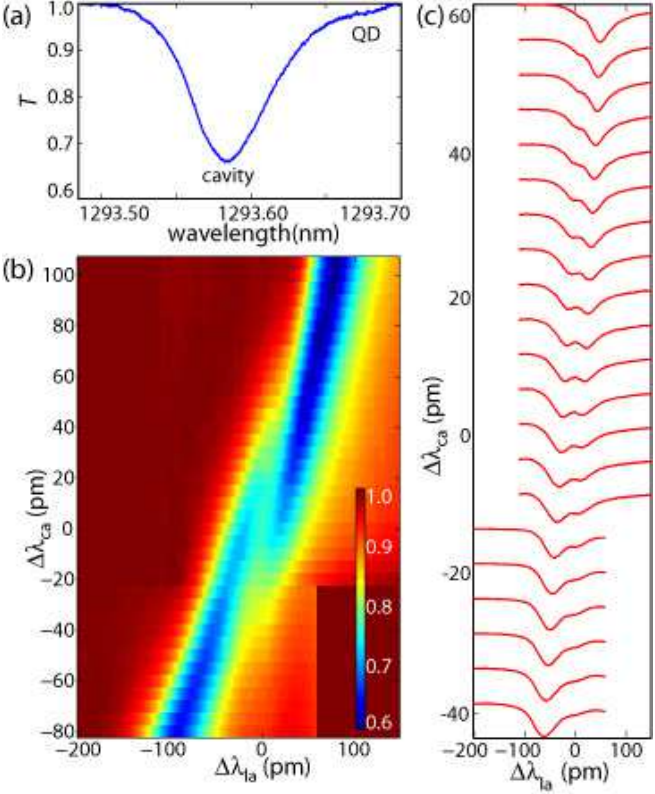}
\caption{(a) Normalized (to unity) transmission spectrum for the
cavity studied in section \ref{sec:bad_cavity_data}, for a
detuning from the QD of $\Delta\lambda_{\text{ca}}=-$100 pm. (b)
Image plot of the transmitted spectra as a function of laser-QD
detuning ($\Delta\lambda_{\text{la}}$) and cavity-QD detuning
($\Delta\lambda_{\text{ca}}$). (c) Series of transmission spectra
for a zoomed-in region of $\Delta\lambda_{\text{ca}}$. Sample
temperature = 8 K} \label{fig:bad_cavity_anticrossing}
\end{center}
\end{figure}

Figure \ref{fig:bad_cavity_anticrossing}(a) shows a transmission
spectrum for the device under consideration, where the cavity mode
is blue-detuned of the QD $X_{a}$ transition by
$\Delta\lambda_{\text{ca}}=-$100 pm, and the system is probed in
the weak driving limit ($n_{\text{cav}}\approx0.01$) at 8 K. In
comparison to the cavity mode studied in the previous section, we
see that this cavity mode has significantly more loss
($Q_{T}=2.1{\times}10^4$), so that the decay rate
$\kappa_{T}/2\pi=$5.4 GHz. In this device, $\kappa_{T}$ dominates
$\gamma_{\beta}$ and as a result, the mode appears as a singlet.
Interpretation of the data is then particularly simple in that the
system behaves very similarly to that of a standard single mode
cavity QED system, with the importance of the additional
complexity of $cw$/$ccw$ modal coupling due to surface roughness
being limited.  Nevertheless, the full QME picture including
$\gamma_{\beta}$ and in particular, the presence of a second
cavity mode, is the most accurate way to study the system, and a
necessity for understanding the presence of a reflected signal,
for example.

Figure \ref{fig:bad_cavity_anticrossing}(b) and
\ref{fig:bad_cavity_anticrossing}(c) show the transmitted spectrum
in the weak driving limit as it is N$_{2}$-tuned towards the QD.
The procedure for tuning the cavity is the same as it was in the
previous section, but the N$_{2}$ flow rate was reduced so that
the step size was 5 pm.  The data show a clear anti-crossing
behavior, with a peak splitting of 37 pm, corresponding to an
approximate coupling strength $g/2\pi\approx3.3$ GHz.  The reduced
sample temperature (8 K) at which the data were taken in
comparison to the measurements in the previous section (15 K)
suggest that the $\gamma_{\perp}$ may also be reduced. Indeed, QME
simulations suggest that the peak width, depth, and contrast of
the transmission spectrum on resonance are best fit by
$\gamma_{\perp}/2\pi=0.57$ GHz, with $\gamma_{p}$ lower by a
factor of 3 relative to its value in Section
\ref{sec:strong_coupling_data} ($\gamma_{\parallel}$ remains
unchanged). With the previously mentioned $\kappa_{T}/2\pi=$5.4
GHz, the system is in the bad-cavity regime, with
$\kappa_{T}>g>\gamma_{\perp}$. Since
$g\gtrsim(\kappa_{T}+\gamma_{\perp})/2$, anti-crossing and
saturation effects seen previously will largely be reproduced
here.

\begin{figure}
\begin{center}
\includegraphics[width=\columnwidth]{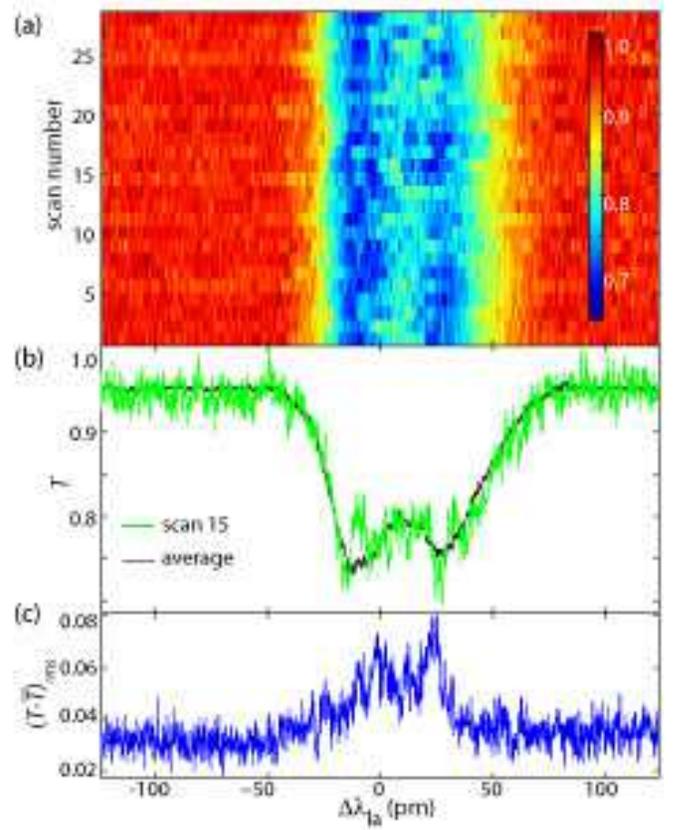}
\caption{(a) Image plot showing multiple un-averaged transmission
spectra, all acquired under the same experimental conditions, for
the device studied in Fig. \ref{fig:bad_cavity_anticrossing} close
to resonance $\Delta\lambda_{\text{ca}}=0$. (b) Plot of scan 15
from (a), shown as a light-colored line, along with the average of
all scans from (a), shown as a dark colored line. (c) Plot of the
ensemble-averaged root-mean-square deviation between the single
scan and averaged transmission data.}
\label{fig:single_vs_avg_scans}
\end{center}
\end{figure}

Before moving on, let us briefly consider an experimental detail
that we have not yet addressed, which is the averaging of the
spectra for each value of $\Delta\lambda_{\text{ca}}$.  This
averaging (typically consisting of 10 to 20 scans) is performed
due to the noisy signal produced in a single scan of the coupled
cavity-QD resonant system (noticeably more noisy than for a
detuned cavity scan).  An example of this is shown in Fig.
\ref{fig:single_vs_avg_scans}, which examines the cavity mode
on-resonance with the neutral exciton line of the QD. We first
show an image plot displaying numerous single scans taken in
succession and under the same resonant optical driving and
detection conditions (analog low-pass filter of 300 Hz and
analog-to-digital sampling time of 1 ms).  This image gives some
indication of the size of the fluctuations present in the single
scan data.  Underneath the image plot, we overlay a representative
single scan with the average of all of the single scans.  We see
that there are pronounced fluctuations in the single scan
transmission data (by as much as 25 $\%$ of the average value)
within the resonance dips.  The source of these fluctuations is
not just electronic noise of our detector, but is related to the
interaction of the cavity with the QD. In particular, the ensemble
averaged root-mean-square (rms) deviation from the mean
transmission signal as a function of probe wavelength, shown in
Fig. \ref{fig:single_vs_avg_scans}(c), clearly increases in the
region of the coupled cavity-QD resonances.  Measurements taken
when the cavity is far detuned from the QD do not show such a
variation for the bare-cavity transmission resonance.  This
suggests that the QD is somehow the source of these increased
fluctuations, perhaps as a result of processes such as spectral
diffusion and blinking, which we are unable to resolve any further
in our current experiments due to the limited time resolution of
our data, and that clearly require further experimentation and
analysis.

\subsection{Nonlinear spectroscopy}

In this section, we consider nonlinear spectroscopy of the system
presented above.  In these measurements, the taper coupling
position with respect to the cavity has been slightly improved, so
that the bare-cavity ${\Delta}T=50$ $\%$ with
$\kappa_{T}/2\pi=4.1$ GHz. We tune the cavity to a position close
to resonance ($\Delta\lambda_{\text{ca}}\approx 0$), and examine
how the spectra change with $n_{\text{cav}}$, both due to
increasing the near-resonant probe beam power and through the
addition of a far off-resonance control beam.  A schematic of the
experimental setup used for these measurements is shown in Fig.
\ref{fig:bad_cavity_pump_probe_beam_saturation}(a).  The control
beam, consisting of laser light from a $1420$ nm to $1500$ nm band
external cavity tunable diode laser, is coupled into the fiber
taper through a directional coupler that combines it with the
$1300$ nm probe beam.  A $1460$ nm to $1500$ nm bandpass filter
(BPF) is used to provide 60 dB of rejection of the spontaneous
emission from the control laser beam in the $1300$ nm wavelength
band prior to input to the fiber taper.  Additionally, in these
measurements a narrowband tunable bandpass filter (TBPF), with a
20 dB full-width at half-maximum bandwidth of $2$ nm, is used to
filter spontaneous emission from the probe laser beam.  At the
fiber taper output, a pair of BPFs providing 120 dB rejection of
the control laser beam are used to remove any residual pump beam
prior to measurement of the low-power $1300$ nm probe beam
transmission.

\subsubsection{Saturation of the Jaynes-Cummings system}

\begin{figure*}
\begin{center}
\includegraphics[width=1.7\columnwidth]{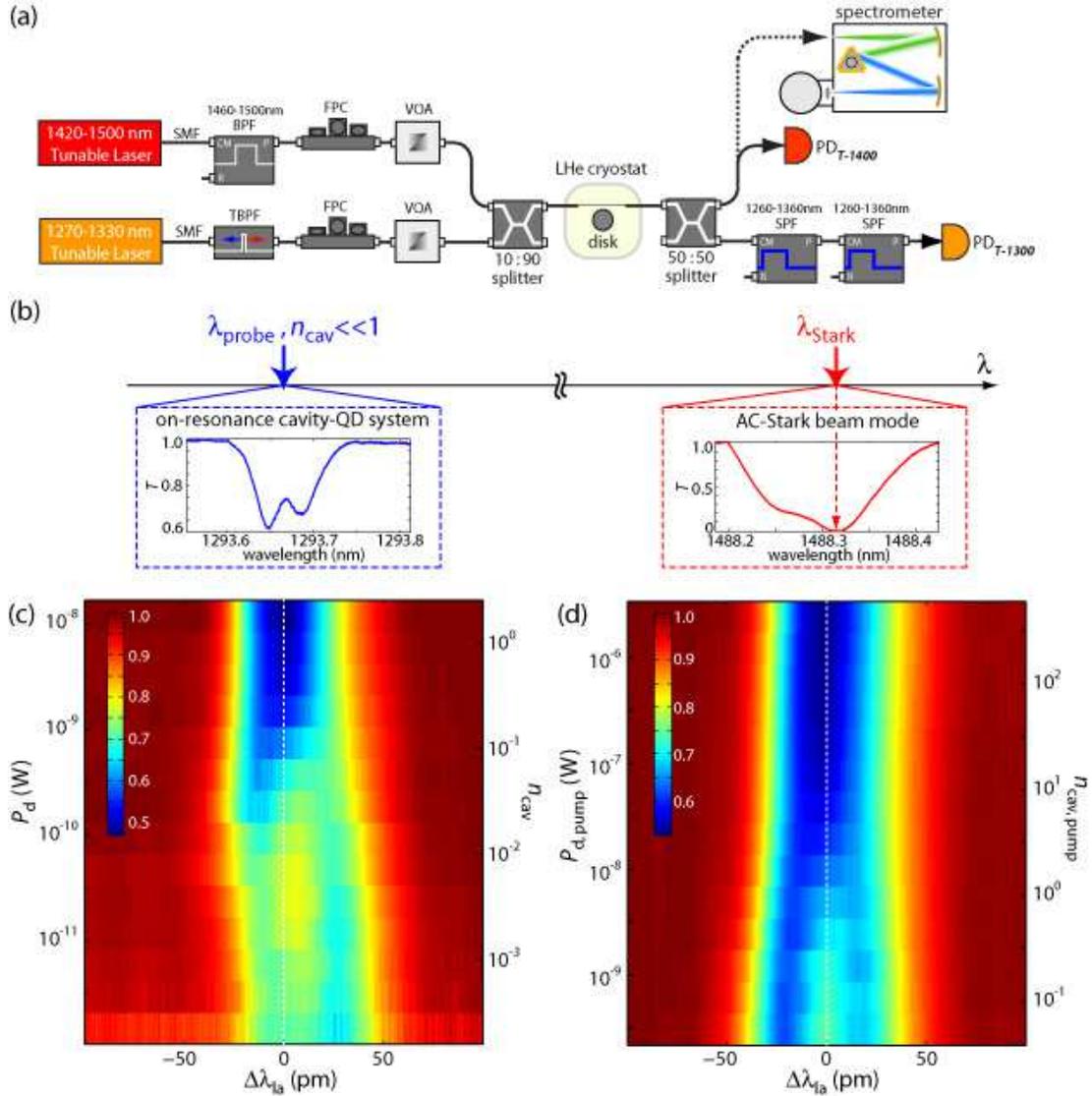}
\caption{(a) Schematic of the experimental setup and procedure
used in pump-probe measurements.  Acronyms for optical components
are: single-mode fiber (SMF), fiber polarization controller (FPC),
variable optical attenuator (VOA), bandpass filter (BPF), tunable
bandpass filter (TBPF), short-pass filter (SPF), photodetector for
the $1300$ nm band probe beam transmission (PD$_{T-1300}$), and
photodetector for the $1400$ nm band ac-Stark control beam
transmission (PD$_{T-1400}$). (b) Transmission scans of the (left)
on-resonance cavity-QD system under weak driving
($n_{\text{cav}}\approx 5{\times}10^{-3}$) and (right) 1488 nm
ac-Stark shift pump beam WGM. (c) Resonant transmission versus
on-resonance probe power. Image plot of the normalized (to unity)
transmission as a function of probe laser-QD detuning
($\Delta\lambda_{\text{la}}$) and probe beam strength (dropped
power into the cavity ($P_{\text{d}}$) on left axis and average
intracavity photon number ($n_{\text{cav}}$) on right axis), for
the device studied in Fig. \ref{fig:bad_cavity_anticrossing}, at a
cavity-QD detuning close to resonance
$\Delta\lambda_{\text{ca}}\approx0$. (d) Resonant transmission
versus ac-Stark pump power.  Image plot of the normalized (to
unity) transmission spectrum as a function of probe laser-QD
detuning ($\Delta\lambda_{\text{la}}$) and pump beam strength
(dropped 1488 nm ac-Stark pump beam power into the cavity
($P_{\text{d,pump}}$) on left axis and average intracavity pump
photon number ($n_{\text{cav,pump}}$) on right axis), at a
cavity-QD detuning close to resonance
$\Delta\lambda_{\text{ca}}\approx0$.}
\label{fig:bad_cavity_pump_probe_beam_saturation}
\end{center}
\end{figure*}

Before considering the effects of the red-detuned pump beam on the
system, we first examine system saturation under increased driving
by the on-resonance probe beam.  Figure
\ref{fig:bad_cavity_pump_probe_beam_saturation}(c) plots the
transmitted spectrum for the system as a function of
$P_{\text{d}}$ and $n_{\text{cav}}$, where each spectrum has been
averaged 10 to 20 times. In contrast with the similar plots in
section \ref{subsec:nonlinear_sc}, here we have plotted the
spectrum as a function of $\Delta\lambda_{\text{la}}$. The section
\ref{subsec:nonlinear_sc} data were plotted against
$\Delta\lambda_{l}$, which was defined relative to the short
wavelength resonance peak/dip in the reflection/transmission
spectrum, a necessity due to a slow fluctuation in the laser
wavelength position during the measurements.  In these
measurements, the laser position was stable enough to avoid the
need to do this. The observed saturation behavior is again very
similar to that seen in the previous section, with the onset
occurring for $n_{\text{cav}}\approx0.1$.

Along with the clear saturation that transforms the doublet
structure under weak driving into a singlet under strong driving,
we see an unexpected feature at the lowest drive powers, where
$n_{\text{cav}}\lesssim0.01$.  At these levels, the spectrum does
not saturate, but does appear to blue shift. Interestingly, the
blue shift does not continue at higher powers. The origin of the
shift is not currently clear, though some immediate possibilities,
such as drift of the center laser frequency, heating, free carrier
generation, and removal of the N$_{2}$ film do not seem to match
the observed behavior.

\subsubsection{Pump-probe spectroscopy, nonlinear emission, and
system saturation} \label{subsec:pump_probe}

We now consider a pump-probe measurement on this coupled
microdisk-QD system.  Pump-probe spectroscopy might be used in a
number of different experiments in the study of this system.  Our
initial interest has been in using an off-resonant beam to
ac-Stark shift the QD while it is coupled to a cavity mode,
providing a level of control of the system. For example, one could
envision using the ac-Stark shift to tune the QD with respect to
the cavity mode, possibly allowing for nondestructive state
readout measurements of the type recently demonstrated in circuit
QED \cite{ref:Schuster}. Another possible application of a
pump-probe measurement would be to take advantage of the ultra-low
power nonlinearity observed in this system for switching
applications, where a pump laser is used to change the
reflection/transmission spectrum of the cavity and its effect on a
probe beam is examined. Finally, pump-probe spectroscopy might be
a means to investigate aspects of the structure of the
Jaynes-Cummings system at a level beyond measurements of vacuum
Rabi splitting and anti-crossing \cite{ref:Thompson2}.

By performing cavity mode spectroscopy in the 1400 nm band, we
find the spectral position of a microdisk WGM at
$\lambda_{\text{pump}}=$1488.32 nm, ensuring that the coupling of
this pump beam into the cavity is efficient. The right inset of
Fig. \ref{fig:bad_cavity_pump_probe_beam_saturation}(b) shows a
wavelength scan of the mode we couple to, which is likely the
TE$_{1,11}$ mode (with the 1300 nm WGM being the TE$_{1,14}$
mode).  From this transmission spectrum, we can accurately
determine the dropped pump power $P_{\text{d,pump}}$ and
intracavity pump photon number $n_{\text{cav,pump}}$.

\begin{figure}
\begin{center}
\includegraphics[width=\columnwidth]{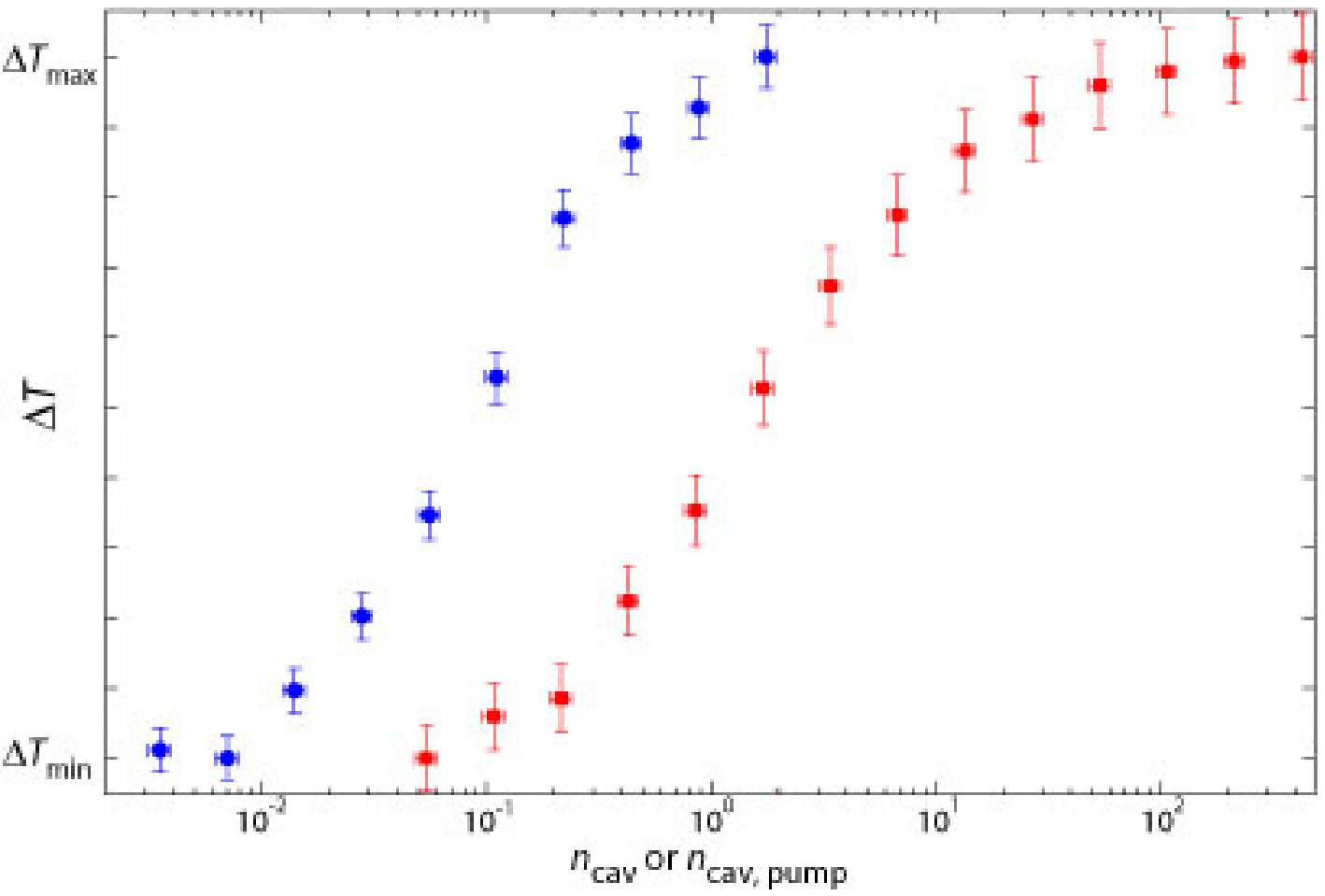}
\caption{Comparison of the two saturation mechanisms discussed for
the device studied in Fig. \ref{fig:bad_cavity_anticrossing}. The
blue circles plot the scaled transmission contrast $\Delta T$
against $n_{\text{cav}}$, where only a probe beam is applied.  The
red squares plot $\Delta T$ against $n_{\text{cav,pump}}$, where
the probe power is fixed at $n_{\text{cav}}\approx0.005$ and the
pump power is varied.  The ordinate axis is scaled such that the
range of transmission contrast is between the minimum and maximum
measured values for each experiment.  The unscaled values of
${\Delta}T_{\text{min}}$ (${\Delta}T_{\text{max}}$) are 0.22
(0.53) for the probe beam only measurement, and 0.26 (0.48) for
the pump-probe measurement.}
\label{fig:bad_cavity_pump_vs_probe_saturation}
\end{center}
\end{figure}

We next fix the 1300 nm probe beam at $P_{\text{d}}=30$ pW,
corresponding to $n_{\text{cav}}\approx5{\times}10^{-3}$, so that
with the 1488 nm pump beam off, we trace out a Rabi-split
transmission spectrum as seen in the left inset of Fig.
\ref{fig:bad_cavity_pump_probe_beam_saturation}(b). We then turn
on the pump laser (at a fixed wavelength
$\lambda_{\text{pump}}=$1488.32 nm) and vary its power while
sweeping the probe beam in frequency and recording the 1300 nm
transmission spectrum. Figure
\ref{fig:bad_cavity_pump_probe_beam_saturation}(d) shows an image
plot of individual spectra from this measurement. We observe a
saturation of the transmission spectrum that, on the surface,
looks very similar to that seen in Fig.
\ref{fig:bad_cavity_pump_probe_beam_saturation}(c), minus the
wavelength blue-shift evident in the probe beam saturation at very
low powers. Although the saturation behavior is quite similar, it
occurs at significantly larger pump beam values than required when
the probe beam was used to saturate the system. Evidence of an
ac-Stark shift of the QD, which would not saturate the spectrum
but rather modify it in manner analogous to that in the
anti-crossing spectrum of Fig. \ref{fig:bad_cavity_anticrossing}
(where now the QD is tuned instead of the cavity), is not seen.

To better compare the two different saturation mechanisms, in Fig.
\ref{fig:bad_cavity_pump_vs_probe_saturation} we examine the
transmission contrast on resonance ($\Delta\lambda_{la}=0$) as a
function of intracavity photon number generated by either the
on-resonance probe beam or the off-resonance ac-Stark shift
control beam. In this data, we have scaled the transmission
contrast from each of the data sets to lie on a common axis of
ordinates (y-axis).  The data show some differences in the shape
of the saturation curves, and most noticeably, that saturation due
to the pump beam occurs for $n_{\text{cav,pump}}\approx1$, nearly
one order of magnitude larger than the $n_{\text{cav}}\approx0.1$
saturation value we see when there is no pump beam and just the
probe beam power is varied.

Naturally, the cause of the saturation effected by the
off-resonant pump beam is an important question.  Recent work
involving similar self-assembled InAs QDs has shown that it is
possible to use relatively intense laser beams to create dressed
states of the bare excitonic levels
\cite{ref:X_Xu,ref:Kroner1,ref:Jundt1,ref:X_Xu2,ref:Gerardot1}. In
Ref. [\onlinecite{ref:Jundt1}] laser instensities as large as $9$
kW/cm$^2$ were used, with no discernable dephasing introduced by
the control laser.  For comparison, we see dephasing effects due
to the $\lambda=1488$ nm control beam at
$n_{\text{cav,pump}}\approx 1$, which corresponds to roughly $5$
kW/cm$^2$ of circulating intensity within the microdisk.  Notable
differences between the system studied here and the QDs studied in
the above mentioned references are the lack of a field-effect
structure \cite{ref:Drexler1} in our device, the close proximity
of dry-etched semiconductor surfaces to the QD in the microdisk,
and the DWELL electronic structure in which the excitonic states
of the QD lie. Embedding the QD within a field-effect structure
may provide a level of stability for the charged environment of
the QD, whereas the presence of nearby surfaces \cite{ref:Michael}
and the DWELL structure \cite{ref:Moreno1} may provide for new
pathways of efficient photoexcitation of free-carriers by the
$\lambda=1488$ nm control laser.  As such, the QD states in our
devices may be more susceptible to spectral diffusion and blinking
\cite{ref:Santori4} due to charge fluctuations in and around the
QD stemming from photoexcited free-carriers.

\begin{figure}
\begin{center}
\includegraphics[width=0.9\columnwidth]{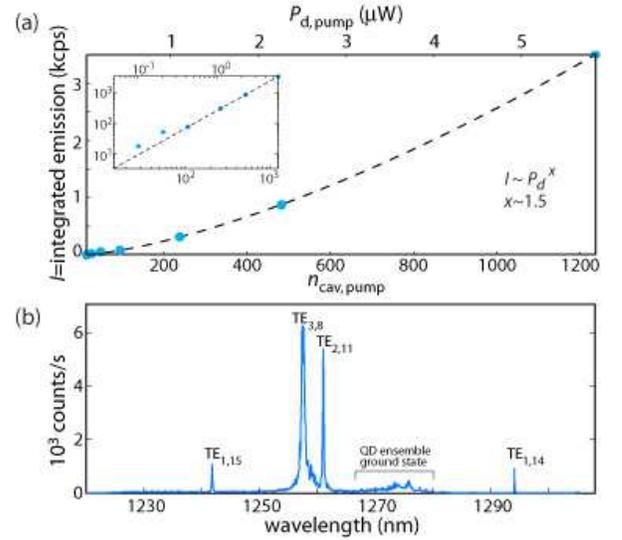}
\caption{(a) Integrated emission ($I$) into the 1293 nm to 1296 nm
wavelength region as a function of $P_{\text{d,pump}}$ and
$n_{\text{cav,pump}}$. A least squares fit assuming $I \propto
(P_{\text{d,pump}})^{x}$ yields $x\approx 1.5$.  The inset shows
the data plotted on a logarithmic scale.  (b) Emission spectrum
over a broad wavelength range for $P_{\text{d,pump}}=10$ $\mu$W.}
\label{fig:1400nm_pump_emission}
\end{center}
\end{figure}

As a first attempt at studying the QD exciton dephasing caused by
the far red-detuned ac-Stark control beam, we turn off the probe
beam and spectrally resolve any emitted light that is generated by
the $1488$ nm control laser beam. Figure
\ref{fig:1400nm_pump_emission}(a) shows the total integrated
emission within a narrow wavelength range ($\lambda=$1293 nm to
1296 nm, covering the cavity mode and QD states $X_{a}$ and
$X_{b}$) as a function of $P_{\text{d,pump}}$. For pump beam
powers as low as $P_{\text{d,pump}}=100$ nW, significant PL is
measured (the smallest pump power levels studied in Fig.
\ref{fig:bad_cavity_pump_probe_beam_saturation}(d) do not produce
detectable emission).  The dependence of emission on pump power is
clearly nonlinear, with a least-squares-fit estimate of $x=1.5$
for the power law exponent.  This would tend to indicate a
nonlinear absorption process is at play here, although
multi-photon absorption in bulk GaAs at a wavelength of $1.5$
$\mu$m is estimated to be negligible \cite{ref:Hurlbut}.  We have
also checked, by monitoring the cavity mode quality factor as a
function of pump power, that any nonlinear absorption is
significantly less than the low-power optical loss of the cavity
mode (exponential absorption coefficient $\alpha=1$ cm$^{-1}$) at
all control laser power levels studied here.

Figure \ref{fig:1400nm_pump_emission}(b) plots the
photoluminescence spectrum over a broad wavelength range, for a
relatively large pump power of $P_{\text{d,pump}}=10$ $\mu$W. We
see a significant amount of emission into other microdisk cavity
modes and into the ensemble of QDs.  Evidently the photogenerated
carriers are free to populate a large portion of the ensemble of
QDs within the microdisk. A possible explanation for the nonlinear
PL intensity dependence, instead of nonlinear absorption, may be a
carrier density dependence of the relaxation and emission process.
The emission process is not well understood for the cavity modes
detuned from excitonic lines of the QDs \cite{ref:Hennessy3}, and
as shown in Fig. \ref{fig:1400nm_pump_emission}(b), cavity mode
emission constitutes a large portion of the PL intensity.
Moreover, linear absorption from deep level defects in the
semiconductor bulk or surface states at the etched edge of the
microdisk has been measured to represent a significant fraction of
the optical loss in similar GaAs microdisks \cite{ref:Michael}.
Further two-wavelength experiments involving tuning of the pump
wavelength and pulsed measurements are being explored to better
understand the physical processes involved.

As a final comment, we compare the saturation mechanism of the
resonant cavity-QD system in the two-wavelength pump-probe
experiment with that of the single-wavelength probe measurements.
In particular, we consider the effect this additional saturation
mechanism would have on measurements of the resonant response of
the system under increasing probe beam power as depicted in Fig.
\ref{fig:power_series_data} and
\ref{fig:bad_cavity_pump_probe_beam_saturation}.  In the
interpretation of these measurements, the saturation was
attributed to the coherent interaction of the probe laser with the
two-level exciton system, and the limited rate at which a single
exciton can spontaneously scatter photons. We now see that, in
addition to this effect, there is the second possibility of
absorption of the probe beam and the effect of generated free
carriers on the QD. However, Figs.
\ref{fig:bad_cavity_pump_probe_beam_saturation} and
\ref{fig:bad_cavity_pump_vs_probe_saturation} show that this
effect is comparatively small until $n_{\text{cav}}\gtrsim0.5$, at
which point the coherent saturation of the system has already
taken strong effect.

\section{Summary}

In summary, we have presented experimental results on the coherent
optical spectroscopy of fiber-coupled semiconductor microdisk-QD
systems.  This coherent spectroscopy is achieved by using a fiber
taper waveguide to efficiently access the semiconductor microcavity,
allowing us to resonantly probe the system and measure its
transmission and reflection spectra, and compare our experimental data
with the results of a quantum master equation model. Along with
measurements of vacuum Rabi splitting and system saturation for less
than one intracavity photon, we consider a pump-probe experiment where
the system response is monitored as a far red-detuned pump beam is
coupled into the cavity and its power is varied.  We find that the
system saturates here as well, at an intracavity pump photon number
that is roughly one order of magnitude larger than the saturation
observed when only the resonant probe beam power is varied. Initial
measurements indicate that sub-bandgap absorption of the pump beam and
subsequent free-carrier generation are the root cause of the
saturation of the resonantly coupled cavity-QD response.  This
suggests that charge fluctuations within and nearby the QD due to
sub-bandgap absorption, even at low excitation powers, may be an
important factor to consider in future pump-probe and nonlinear
spectroscopy experiments of the DWELL-microcavity system.  Further
studies involving variation of the control laser wavelength,
incorporation of bias gates for charge-control of the QD, and
treatments of the etched microcavity device surface are planned to
better understand and tame this parasitic effect.

\section{Acknowledgements}

We thank A. Stintz and S. Krishna of the Center for High
Technology Materials at the University of New Mexico for providing
the quantum dot material growth. This work was partly supported by
the DARPA NACHOS program.

%\subsubsection{Prospects for AC-Stark shifting of the QD using a far
%off-resonant cavity mode}
\bibliography{./PBG_5_7_2008}

\end{document}